\documentclass[a4paper,preprint,12pt]{elsarticle}

\usepackage[T1]{fontenc}
\usepackage{amsmath,amssymb,bm,graphicx,booktabs,microtype}
\usepackage[hidelinks]{hyperref}

\graphicspath{{figures/}}

\newcommand{\W}{\bm W}
\newcommand{\vv}{\bm u}
\newcommand{\xx}{\bm x}
\newcommand{\nn}{\bm n}
\newcommand{\av}[1]{\overline{#1}}
\newcommand{\dd}{\,\mathrm d}
\newcommand{\eqnref}[1]{Eq.~\eqref{#1}}

\newcommand{\panel}[3]{%
	\begin{minipage}[t]{#1\linewidth}
		\centering
		\includegraphics[width=\linewidth]{#2.png}\par
		\vspace{2pt}{\small #3\par}
	\end{minipage}%
}

\begin{document}
	\begin{frontmatter}
		\title{Mesh Sensitivity of Different Boundary Treatments in High-Order Compact Gas-Kinetic Scheme for Hypersonic Heat-Flux Prediction}
		
		\author[xjtu]{Xingyi Jiang}
		
		\author[hkust]{Jiawang Zhang}
		\author[hkust]{Junlei Mu}
		
		\author[xjtu_lab]{Xing Ji\corref{cor1}}
		\cortext[cor1]{Corresponding author. Email addresses:
			\texttt{2236525849@stu.xjtu.edu.cn} (Xingyi Jiang);
			\texttt{jixing@xjtu.edu.cn} (Xing Ji).}
		
		\address[xjtu]{School of Aerospace Engineering, Xi'an Jiaotong University, Xi'an, China}
		
		\address[hkust]{Department of Mathematics, Hong Kong University of Science and Technology, Clear Water Bay, Kowloon, Hong Kong}
		
		\address[xjtu_lab]{State Key Laboratory for Strength and Vibration of Mechanical Structures, Shaanxi Key Laboratory of Environment and Control for Flight Vehicle, School of Aerospace Engineering, Xi'an Jiaotong University, Xi'an, China}
		
		\begin{abstract}
			Accurate wall-heat-flux prediction in hypersonic flow depends on both the wall treatment and the available near-wall resolution. This work compares three complete implementations of a gas-kinetic scheme (GKS) for Mach~8.03 laminar flow over an isothermal circular cylinder: a linear Green--Gauss scheme (GG), a compact ghost-cell scheme (GC), and a pointwise one-sided compact scheme (POS). In POS, the stationary isothermal-wall conditions are imposed at the two Gaussian points used by the kinetic wall-flux quadrature. A reference mesh and six meshes in the uniform reduction (UR), log-layer unpacked (LLU), and log-layer packed (LLP) families are used to assess heat-flux sensitivity. At 24 digitized experimental angles, GG has the lowest root-mean-square difference on the reference, UR, and LLP families, whereas POS is closest on the two LLU meshes; GC is not the lowest-RMS scheme on any tested mesh. Hence no treatment is uniformly superior. On the present high-aspect-ratio cylinder grids, GG is the more robust choice except under LLU node redistribution, for which POS is preferable. Uniform reduction produces the largest heat-flux response for all three implementations. The comparison concerns complete schemes and does not isolate the pointwise constraints from differences in reconstruction and geometry representation.
		\end{abstract}
		\begin{keyword}
			Gas-kinetic scheme \sep Curved wall \sep Compact reconstruction \sep
			Hypersonic heat transfer \sep Wall-normal mesh resolution
		\end{keyword}
	\end{frontmatter}
	
	\section{Introduction}
	
	Aerodynamic heating is a central design constraint for hypersonic vehicles.  Strong compression and viscous transport near a blunt surface produce high wall temperatures and heat loads; at sufficiently high enthalpy, thermochemical and gas--surface rate processes add further complexity \cite{fayriddell1958,candler2019}.  Even in the calorically perfect, continuum model considered here, wall heat flux is controlled by the near-wall thermal state and its normal temperature variation.  Accurate heat-flux prediction therefore requires both a consistent wall treatment and sufficient resolution of the near-wall thermal layer.
	
	For a stationary isothermal wall, the continuum conditions are no slip and prescribed temperature: $\vv=\bm0$ and $T=T_w$.  Their discrete imposition is not routine in a high-order method because the interior stencil is truncated at the boundary while both solution values and viscous derivatives contribute to the flux.  High-order finite-difference methods may impose the conditions weakly through penalty terms or construct boundary and ghost-point data by inverse Lax--Wendroff procedures \cite{svard2008,liu2023}.  Finite-element formulations can impose Dirichlet data weakly in the variational form, as in Nitsche's method \cite{nitsche1971}.  In cell-centred finite-volume and immersed-boundary implementations, missing exterior or interface information is commonly supplied by an auxiliary reconstruction or interpolation from the fluid side \cite{depalma2006}.  Across these approaches, curved geometry, stretched near-wall cells, stability, conservation, and preservation of the intended interior order must be addressed together; prescribing the wall value alone does not supply all data required by a high-order viscous flux.
	
	A gas-kinetic scheme (GKS) obtains the interface flux from the evolution of a gas distribution function, so reconstructed states and their gradients enter the wall flux \cite{xu2001}.  Compact versions of the GKS additionally use updated cell-averaged gradients in their reconstruction, which makes the wall closure part of the reconstruction rather than a post-processing condition \cite{pan2016,ji2020,zhao2023,ji2024}.  Zhang, Ji and Xu \cite{zhang2026} constructed a one-sided third-order compact boundary treatment that incorporates no-slip-wall information and a normal-pressure-gradient approximation directly into the compact reconstruction, avoiding a ghost cell at a curved boundary.  Their work also makes clear that the Green--Gauss, ghost-cell compact, and one-sided compact choices are distinct complete scheme implementations, not merely different algebraic forms of a single boundary value.
	
	The present study assesses how these three wall treatments affect predicted heat flux on the Mach~8.03 isothermal-cylinder case.  GG uses a linear Green--Gauss reconstruction, GC uses a compact quadratic reconstruction with ghost-cell wall data, and POS uses a compact one-sided quadratic reconstruction on quadratic wall geometry.  Unlike GC, which introduces wall information through an auxiliary ghost cell, POS places the wall equations directly at the two Gaussian points used by the kinetic wall-flux quadrature.  The POS candidate is therefore constrained at the actual flux locations rather than only through a face-averaged wall condition.  The aim is not to claim an isolated effect of these point constraints, since reconstruction, geometry representation, and pressure output also differ among the complete schemes.  Instead, the comparison identifies which complete wall-treatment implementation gives the closer heat-flux prediction for each of the specified near-wall mesh families.
	
	The wall-normal test modifies a fixed near-wall band in three different ways.  UR reduces the interval count throughout the band; LLU redistributes the nodes while retaining the interval count; and LLP merges only the innermost intervals while retaining the outer reference nodes.  Each modification is used at two levels.  The designs are used to test the sensitivity of the three wall-treatment implementations to the distribution of near-wall resolution; they do not constitute a global grid-refinement study or establish an asymptotic order of convergence.
	
	Section~\ref{sec:framework} gives the kinetic ingredients and scheme differences needed to interpret the comparison.  Section~\ref{sec:pointwise} describes the POS wall construction and its limitations.  Section~\ref{sec:case} defines the cylinder case and the mesh modifications.  Section~\ref{sec:results} compares the heat-flux predictions with the digitized experimental points, then treats reference-relative mesh sensitivity, the reported pressure response, and temperature profiles derived from stored cell averages.

	\section{Gas-kinetic framework and comparison schemes}
	\label{sec:framework}
	
	\subsection{Kinetic evolution and finite volume transport}
	Consider a two-dimensional calorically perfect gas with conservative variables
	\begin{equation}
		\begin{gathered}\W=(\rho,\rho U,\rho V,\rho E)^{\mathsf T},\\
			p=(\gamma-1)\left[\rho E-\tfrac12\rho(U^2+V^2)\right].
		\end{gathered}\label{eq:state}
	\end{equation}
	The specific energy is $E=C_vT+(U^2+V^2)/2$, with $C_v=R_g/(\gamma-1)$ and $p=\rho R_gT$. The model describes laminar continuum flow with constant specific heats. Reconstructed states used in the kinetic solver require $\rho>0$ and $p>0$; the equality constraints introduced below do not by themselves guarantee these inequalities.
	
	The BGK evolution model \cite{bgk1954} and equilibrium distribution are
	\begin{align}
		f_t+\vv\cdot\nabla_{\xx}f &= (g-f)/\tau,\label{eq:bgk}\\
		g &= \rho(\lambda/\pi)^{(K+2)/2}
		e^{-\lambda[(u-U)^2+(v-V)^2+|\bm{\xi}|^2]} .\label{eq:maxwell}
	\end{align}
	Here $\vv=(u,v)$ is molecular velocity, $\bm{\xi}\in\mathbb R^K$ represents the remaining molecular degrees of freedom, $\lambda=1/(2R_gT)$, and $K=(4-2\gamma)/(\gamma-1)$. Thus $K=3$ for $\gamma=1.4$ in this two-dimensional representation. Define
	\begin{equation}
		\begin{gathered}
			\bm\psi=\left(1,u,v,\tfrac12(u^2+v^2+|\bm{\xi}|^2)\right)^{\mathsf T},\\
			\dd\Xi=\dd u\dd v\dd\bm{\xi} .
		\end{gathered}
	\end{equation}
	Compatibility requires $\int(g-f)\bm\psi\dd\Xi=0$. The macroscopic state and outward-normal flux follow from
	\begin{equation}
		\begin{aligned}
			\W       &=\int f\bm\psi\dd\Xi,\\
			\bm F_n &=\int(\vv\cdot\nn)f\bm\psi\dd\Xi.
		\end{aligned}
		\label{eq:moments}
	\end{equation}
	
	For a cell $\Omega_i$, conservation gives
	\begin{equation}
		\frac{\dd\av\W_i}{\dd t}=\mathcal L_i(t)
		=-\frac{1}{|\Omega_i|}\sum_{f\subset\partial\Omega_i}
		\int_f\bm F_n(\xx,t)\dd s .\label{eq:fv}
	\end{equation}
	The interface distribution is constructed from the characteristic integral solution of \eqnref{eq:bgk},
	\begin{equation}
		\begin{aligned}
			f(\xx_G,t)={}&e^{-t/\tau}f_0(\xx_G-\vv t)\\
			&+\frac1\tau\int_0^t e^{-(t-t')/\tau}
			g(\xx_G-\vv(t-t'),t')\dd t'.
		\end{aligned}
		\label{eq:integral}
	\end{equation}
	Here $t$ is measured from the beginning of the current step, velocity arguments are suppressed, and $\tau$ is frozen during the local evolution. Reconstructed states and spatial derivatives determine the left and right initial distributions. For each local Maxwellian, define the velocity-dependent coefficients $\bm a=g^{-1}\nabla_{\xx}g$ and $A=g^{-1}\partial_tg$. The first-order Chapman--Enskog distribution is
	\begin{equation}
		\begin{gathered}
			f_{\rm NS}=g[1-\tau_{\rm phy}(\bm a\cdot\vv+A)],\\
			\int g(\bm a\cdot\vv+A)\bm\psi\dd\Xi=\bm0.
		\end{gathered}
		\label{eq:ce}
	\end{equation}
	The second relation determines the temporal coefficient from compatibility. The non-equilibrium term supplies viscous transport; the spatial and temporal expansion of the interface equilibrium supplies the evolution contribution \cite{zhang2026}. This construction explains why both values and gradients from the wall reconstruction enter the flux. The physical relaxation time is $\tau_{\rm phy}=\mu(T)/p$, with Sutherland viscosity. For shock capturing, the viscous implementation uses
	\begin{equation}
		\tau_{\rm num}=\tau_{\rm phy}
		+C\frac{|p_L-p_R|}{p_L+p_R}\Delta t .
	\end{equation}
	Here $C=5.0$ in the supplied cylinder configurations and $p_L,p_R$ are the reconstructed interface pressures. The present calculation also uses the baseline numerical term $0.1\Delta t$ in inviscid regions. The physical relaxation time remains in the non-equilibrium correction, while numerical relaxation controls the decay in the interface model. The non-unit Prandtl number is imposed by the energy-flux correction described in \cite{xu2001}.
	
	\subsection{Time advancement and evolved gradients}
	Let $\mathcal F(\delta)=\int_0^\delta\bm F_n(t)\dd t$. A linear-in-time surrogate for the flux is determined by matching its full- and half-step integrals. Its coefficients, denoted $\bm F_n^n$ and $\partial_t\bm F_n^n$, are
	\begin{align}
		\bm F_n^n&=\frac{4\mathcal F(\Delta t/2)-\mathcal F(\Delta t)}{\Delta t},\nonumber\\
		\partial_t\bm F_n^n&=\frac{4[\mathcal F(\Delta t)-2\mathcal F(\Delta t/2)]}{\Delta t^2}.
		\label{eq:fluxcoef}
	\end{align}
	The present steady-flow implementation employs a single-stage second-order update,
	\begin{equation}
		\av\W_i^{n+1}=\av\W_i^n+\Delta t\mathcal L_i^n
		+\tfrac12\Delta t^2\partial_t\mathcal L_i^n .\label{eq:time}
	\end{equation}
	The time step used for the viscous calculation is
	\begin{equation}
		\Delta t=\mathrm{CFL}\min_i\left\{
		\frac{\Delta h_i}{|\bm U_i|+c_i},
		\frac{\rho_i\Delta h_i^2}{4\mu_\infty}
		\right\},
		\label{eq:timestep}
	\end{equation}
	where $\Delta h_i$ is the cell characteristic length, $c_i$ is the
	local sound speed and $\mu_\infty=\mu(T_\infty)$. The cylinder
	configurations use $\mathrm{CFL}=0.1$.
	
	For the compact schemes, the divergence theorem relates the cell-averaged gradient to the boundary trace:
	\begin{equation}
		\av{\nabla\W}_i^{n+1}
		=\frac1{|\Omega_i|}\sum_f\int_f\W(\xx,t^{n+1})\otimes\nn\dd s .
		\label{eq:gradient}
	\end{equation}
	At interior faces, moments of the evolving interface distribution provide this trace. At solid walls, the boundary solver supplies an auxiliary state for the gradient update; it must be specified separately from the half-space flux construction. These interface states, together with the cell averages, supply the reconstruction data for the compact schemes.
	
	For a curved face parameterized by $\xx(s)$, $s\in[0,1]$, the POS
	implementation accumulates flux and gradient contributions using the
	same Gaussian points and arc-length factors. With
	$j_f(s)=|\xx'(s)|$, the two face integrals are approximated by
	\begin{align}
		\int_f\bm F_n\,\dd s
		&\simeq \sum_{k=1}^{2}\omega_k j_f(s_k)\bm F_{n,k},
		\nonumber\\
		\int_f\W\otimes\nn\,\dd s
		&\simeq \sum_{k=1}^{2}\omega_k j_f(s_k)
		\W_k\otimes\nn_k .
		\label{eq:curved-quadrature}
	\end{align}
	Here $\omega_1=\omega_2=1/2$, $\nn_k$ is the local face normal,
	and $\W_k$ is the evolved trace used for the gradient update.
	The same quadrature structure is used on straight faces with
	constant arc-length factor and normal.
	
	\subsection{Compact reconstruction and comparison schemes}
	An affine local coordinate system $\xx=\xx_c+\bm J\bm r$, with $\bm r=(\xi,\eta)^{\mathsf T}$, scales the reconstruction stencil. The columns of $\bm J$ are two cell-edge vectors. Gradients transform as
	\begin{equation}
		\nabla_{\bm r}Q=\bm J^{\mathsf T}\nabla_{\xx}Q,
		\qquad \nabla_{\xx}Q=\bm J^{-\mathsf T}\nabla_{\bm r}Q .
		\label{eq:transform}
	\end{equation}
	This affine scaling is separate from the isoparametric description of a curved boundary. With the origin at the cell centroid, define
	\begin{equation}
		\begin{gathered}
			\bm\phi=(\xi,\eta,\phi_3,\phi_4,\phi_5)^{\mathsf T},\\
			\phi_3=\xi^2-\av{\xi^2}_i,\quad
			\phi_4=\xi\eta-\av{\xi\eta}_i,\\
			\phi_5=\eta^2-\av{\eta^2}_i.
		\end{gathered}
	\end{equation}
	A scalar conservative component is reconstructed as
	\begin{equation}
		P_Q(\xx)=\av Q_i+\bm\phi(\xx)^{\mathsf T}\bm a_Q,
		\qquad \bm a_Q\in\mathbb R^5 .\label{eq:poly}
	\end{equation}
	An overbar denotes a physical cell average, evaluated with the same geometry and quadrature as the cell centroid and moments. Consequently $\av\xi_i=\av\eta_i=0$ and $\av{\bm\phi}_i=\bm0$, so the central-cell average is preserved. Neighboring cell averages supply equality conditions, and neighboring average gradients supply the least-squares objective \cite{zhang2026}. A quadratic candidate provides the basis for third-order spatial approximation; the achieved order also depends on geometry, flux integration, conditioning and nonlinear selection.

	\subsection{Comparison schemes}
	
	The three schemes are denoted GG, GC and POS. They use the same kinetic transport framework and differ in the spatial information supplied to it. GG uses the linear Green--Gauss reconstruction throughout the computational domain on straight-sided cells. For a scalar component, its uncompressed gradient is
	\begin{equation}
		\bm G_i^{\rm GG}
		=\frac{1}{|\Omega_i|}
		\sum_{f\subset\partial\Omega_i}
		\ell_f\,\frac{\av Q_i+\av Q_{j(f)}}{2}\,\nn_f .
		\label{eq:gg-gradient}
	\end{equation}
	For an interior face $j(f)$ is the adjacent fluid cell. At the wall, its data are supplied by the wall auxiliary state. The GG procedure scales the gradient by the cell feedback factor and evaluates the linear polynomial $\av Q_i+\alpha_i\bm G_i^{\rm GG}\cdot(\xx-\xx_i)$ at the face midpoint. The linear term has zero cell mean for a centroid origin. Thus GG is not a wall-only reference construction; its reconstruction is linear throughout the domain.
	
	GC uses quadratic compact reconstruction on the same straight-sided geometry. At an interior cell, neighboring averages constrain the candidate while neighboring cell-averaged gradients enter the least-squares fit. At a wall cell, the stencil includes an auxiliary ghost cell. The isothermal boundary routine preserves the input pressure when constructing this auxiliary state, assigns zero velocity and temperature $T_w$, and forms conservative derivatives after setting the normal pressure derivative to zero in the auxiliary gradient construction. Thus both values and gradients from the wall side enter the compact fit. The kinetic face flux is sampled at two Gaussian points.
	
	POS uses a compact quadratic reconstruction on quadratic geometry. At a wall cell the fluid-side neighbors are retained, and the auxiliary ghost-cell data in the quadratic wall stencil are replaced by the constraints at the wall quadrature points described in Section~\ref{sec:pointwise}. Wall fluxes are again sampled at two points. The differences between the three schemes therefore include the form of the spatial reconstruction and, for POS, the geometry representation.
	
	\begin{table}[!tb]
		\centering
		\caption{Settings relevant to the comparison of the three complete schemes. All cases use the same gas model, free-stream and wall conditions, and kinetic flux framework.}
		\label{tab:scheme-settings}
		\footnotesize
		\begin{tabular}{p{.08\linewidth}p{.20\linewidth}p{.22\linewidth}p{.16\linewidth}p{.18\linewidth}}
			\toprule
			Scheme & Reconstruction and geometry & Wall information in reconstruction & Wall-flux sampling & Reported pressure diagnostic \\
			\midrule
			GG & Linear Green--Gauss throughout; straight-sided cells & Auxiliary wall state in the Green--Gauss gradient & Face midpoint & Stored interface state \\
			GC & Compact quadratic; straight-sided cells & Auxiliary ghost-cell values and gradients & Two Gaussian points & Wall-adjacent cell-average state \\
			POS & Compact quadratic; quadratic wall geometry & Two pointwise wall constraints on the one-sided quadratic candidate & Two Gaussian points & Arc-length-weighted fluid-side values at the two Gaussian points \\
			\bottomrule
		\end{tabular}
	\end{table}
	
	For POS, the implemented sequence is: construct the constrained quadratic candidates for momentum and total energy; apply nonlinear selection; obtain the two density targets from the selected energy and $T_w$; construct and select the density candidate; evaluate the fluid-side state and gradients at each Gaussian point; and finally compute the kinetic wall flux. The constraints are therefore associated with the reconstruction candidate, whereas the reported flux uses the state retained by the selection procedure.
	
	% Figure 1
	\begin{figure}[!tb]
		\centering
		\panel{.31}{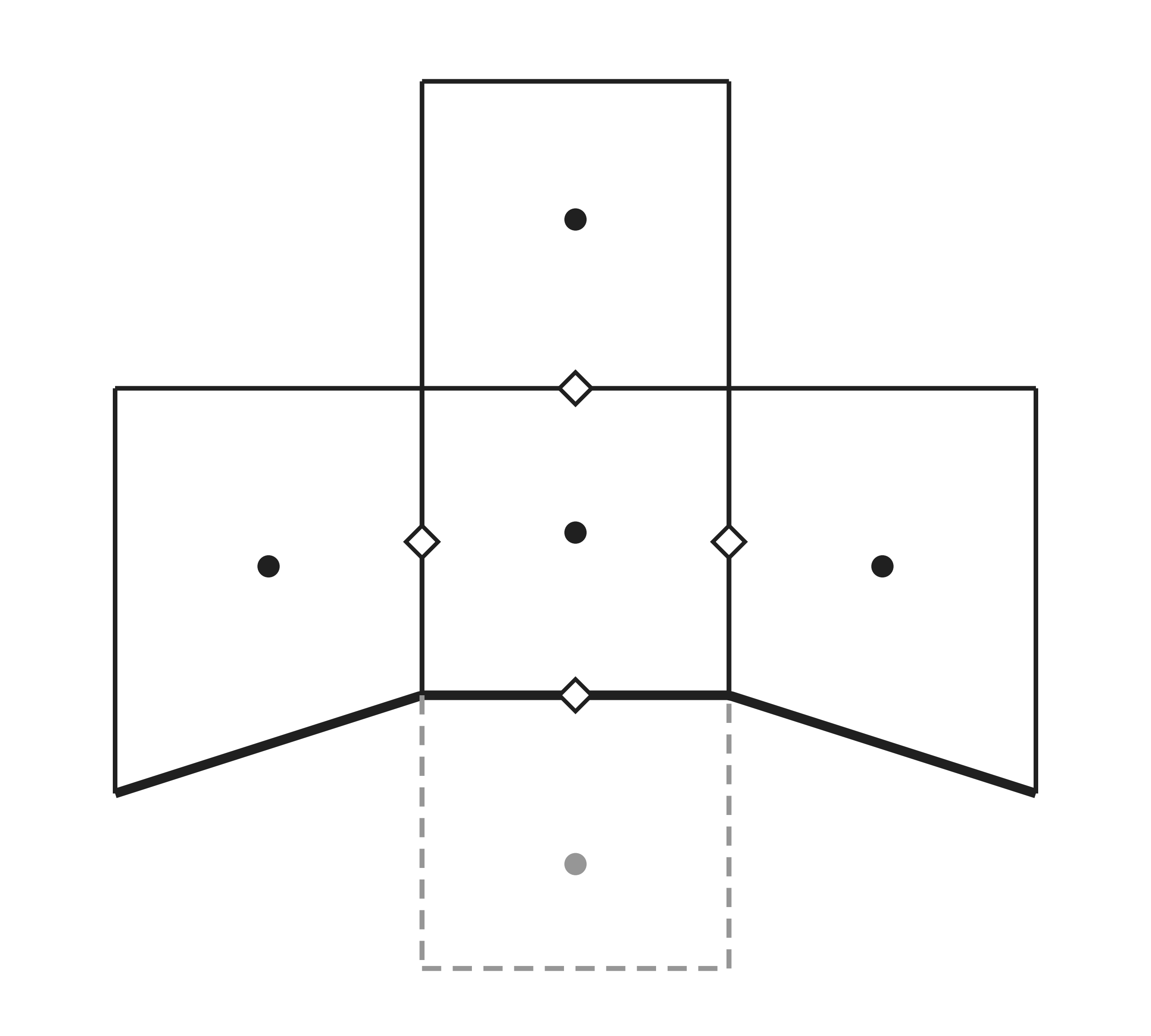}{(a) Green--Gauss\\(GG)}\hfill
		\panel{.31}{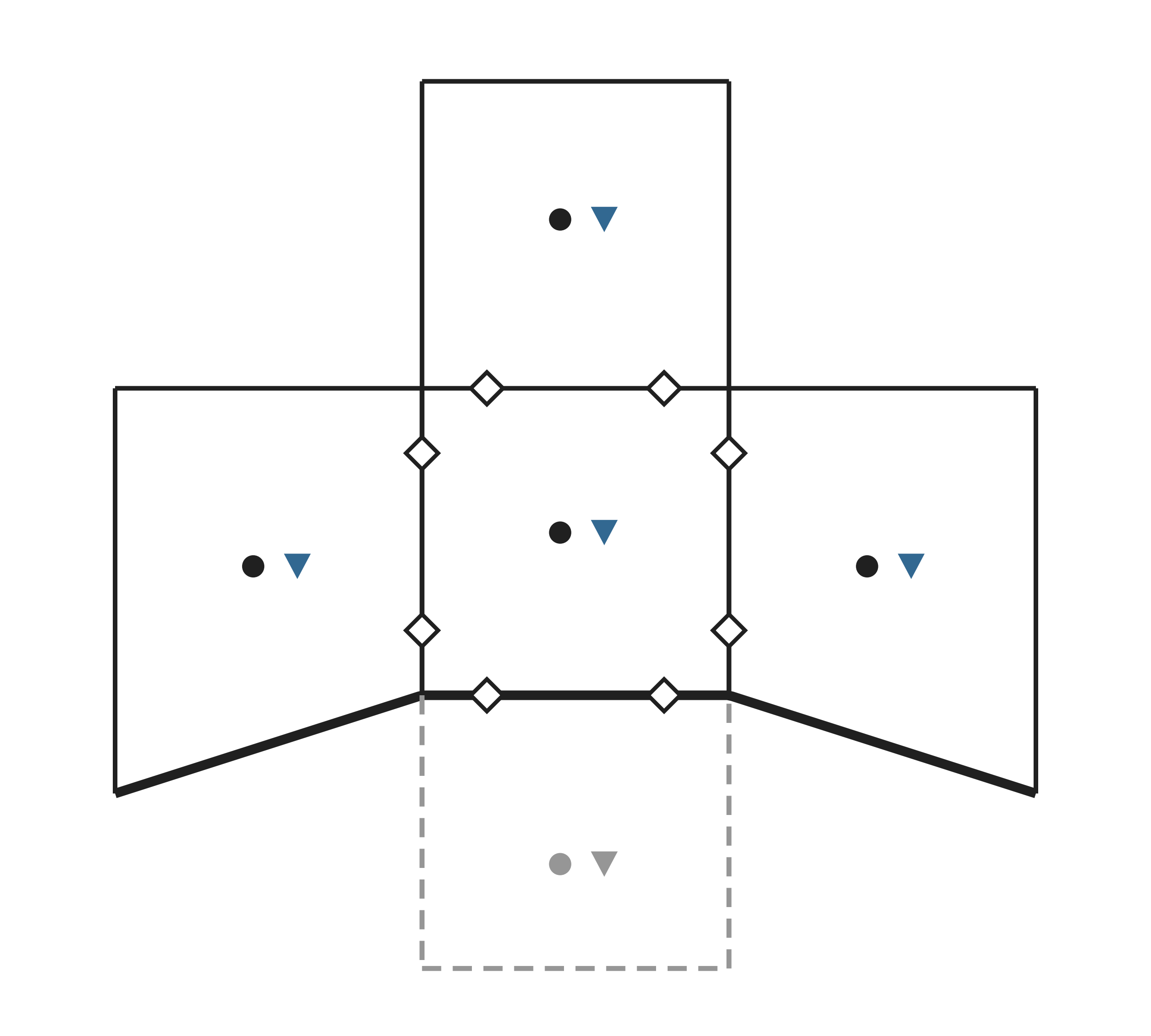}{(b) Compact ghost cell (GC)}\hfill
		\panel{.31}{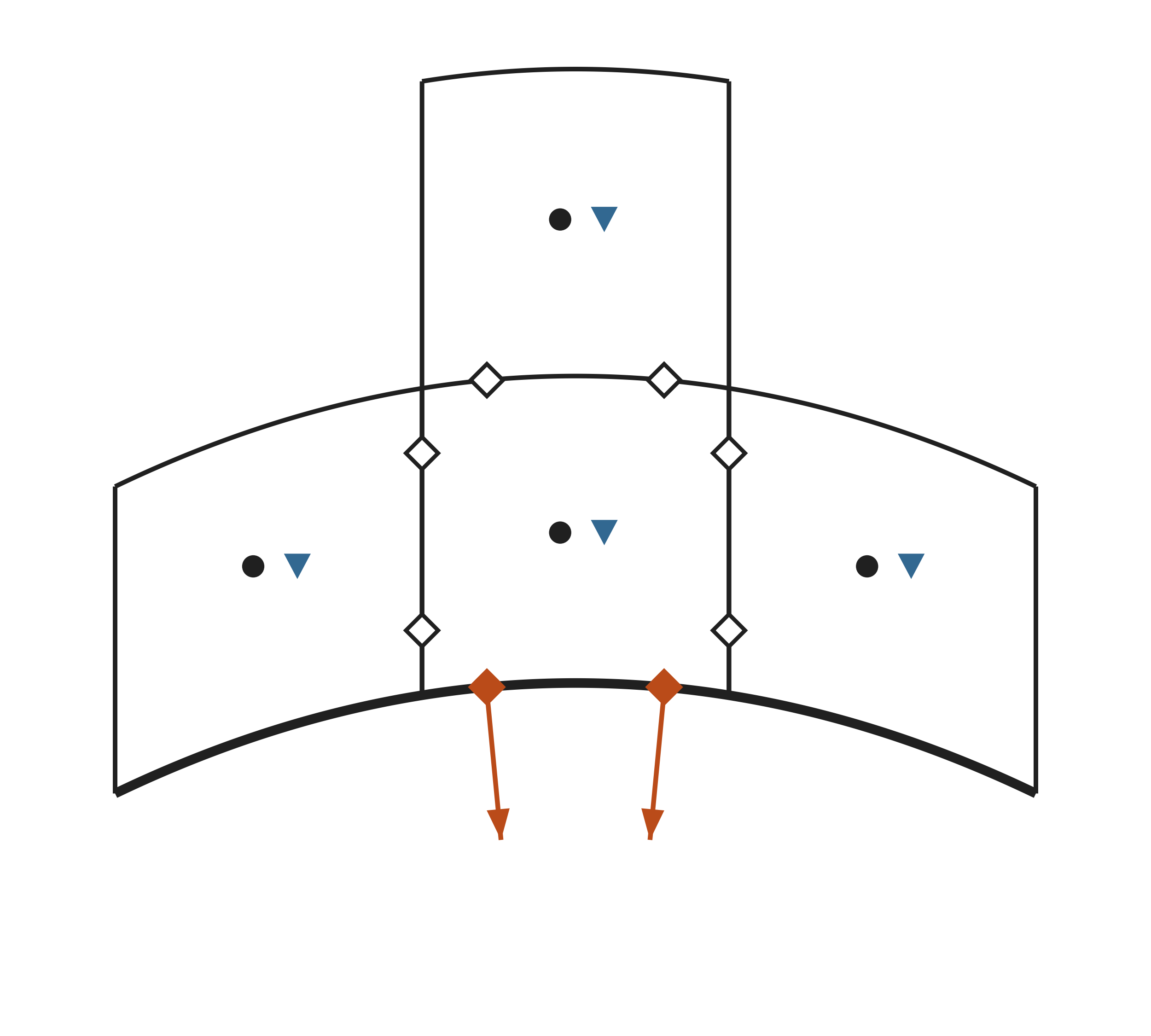}{(c) Compact one-sided (POS)}
		\caption{Wall reconstruction stencils. Circles mark cells supplying averages and triangles indicate available cell-averaged gradients. The marks denote cell data. Diamonds indicate face flux samples, dashed edges delineate the auxiliary cells, and the heavy edge marks the wall. In (c), the two colored diamonds are the constrained Gaussian points and the arrows show their outward local normals.}
		\label{fig:stencil}
	\end{figure}

	\section{Pointwise wall-constrained reconstruction}
	\label{sec:pointwise}
	
	\subsection{Isothermal boundary conditions and curved geometry}
	
	For a stationary isothermal wall, the velocity is zero and the temperature is prescribed as $T_w$. Let $\mathcal E=\rho E$. Differentiating the equation of state at a smooth no-slip wall gives
	\begin{equation}
		\mathcal E_w=\frac{p_w}{\gamma-1},
		\qquad
		\left.\partial_n\mathcal E\right|_w
		=\frac{1}{\gamma-1}\left.\partial_n p\right|_w .
		\label{eq:wallenergy}
	\end{equation}
	The derivative of the kinetic-energy density vanishes there because it is quadratic in velocity. The reconstruction uses the additional boundary-layer approximation $\partial_n p\simeq0$ adopted by Zhang et al.\ \cite{zhang2026}. Its validity depends on the local normal pressure variation. Under this approximation, the normal derivative of total-energy density is set to zero. From $p=\rho R_g T$, one then obtains at the wall
	\begin{equation}
		T_w\partial_n\rho+\rho_w\partial_n T\simeq 0,
		\qquad
		q_n=-\kappa\partial_n T
		\simeq\kappa\frac{T_w}{\rho_w}\partial_n\rho ,
		\label{eq:heatgradient}
	\end{equation}
	where $\kappa=\mu C_p/\mathrm{Pr}$ and the unit normal points out of the fluid. Thus a zero normal derivative of total-energy density still allows conductive heating through a finite temperature gradient. Equation~\eqref{eq:heatgradient} is a continuum relation. The energy flux reported below is computed by the kinetic wall solver.
	
	The curved wall segment is interpolated through its end nodes $\xx_0$, $\xx_1$ and its mid-edge node $\xx_m$,
	\begin{equation}
		\xx(s)=(1-3s+2s^2)\xx_0+(-s+2s^2)\xx_1
		+(4s-4s^2)\xx_m,\qquad s\in[0,1].
		\label{eq:curve}
	\end{equation}
	The two Gaussian parameters are $s_{1,2}=(1\mp1/\sqrt3)/2$. Write $\xx_{G,k}=\xx(s_k)$, $j_s(s)=|\xx'(s)|$ and $\nn_k=\nn(s_k)$. The normal points into the cylinder, and its direction is evaluated separately at each point. The same two locations are used for wall flux sampling. Figure~\ref{fig:stencil} shows the data used by the three wall stencils.
	
	\subsection{What the two constraints determine}
	
	For the unlimited quadratic momentum and energy candidates, POS imposes
	\begin{align}
		P_{\rho U}(\xx_{G,k})&=P_{\rho V}(\xx_{G,k})=0,
		\nonumber\\
		\nn_k^{\mathsf T}\nabla_{\xx}P_{\mathcal E}(\xx_{G,k})&=0,
		\qquad k=1,2.
		\label{eq:pointwall}
	\end{align}
	This choice supplies a boundary condition at each of the two locations where the kinetic solver needs the fluid-side state and gradients. A line-averaged condition, as used by Zhang et al.\ \cite{zhang2026}, supplies one scalar equation per variable and wall face. To distinguish the information in the two constructions, let $r(s)$ be the residual of a scalar value or normal-derivative condition. The exact line average and its normalized two-point approximation are
	\begin{align}
		\langle r\rangle_f
		&=\frac{\int_0^1r(s)j_s(s)\dd s}{\int_0^1j_s(s)\dd s},
		\nonumber\\
		\langle r\rangle_{f,h}
		&=\alpha_1r_1+\alpha_2r_2,
		\qquad \alpha_k=\frac{\omega_k j_s(s_k)}
		{\sum_{j=1}^{2}\omega_j j_s(s_j)},\qquad r_k=r(s_k),
		\label{eq:average}
	\end{align}
	with $\omega_k=1/2$. The relation between the discrete mean and point values is
	\begin{equation}
		r_1=r_2=0
		\quad\Longleftrightarrow\quad
		\langle r\rangle_{f,h}=0,\qquad r_2-r_1=0 .
		\label{eq:pos-mean-difference}
	\end{equation}
	The transformation is invertible because $\alpha_1+\alpha_2=1$. The mean can vanish through cancellation of residuals at the two points. The second equation removes this cancellation and therefore adds a scalar condition. Equation~\eqref{eq:average} normalizes the arc-length weights to form a mean. Equation~\eqref{eq:curved-quadrature} uses the unnormalized weights to form face integrals.
	
	Two vanishing samples still leave part of the variation between them undetermined. On a straight face, the quadratic value residual
	\begin{equation}
		r(s)=\left(s-\tfrac12\right)^2-\tfrac1{12}
		\label{eq:trace-example}
	\end{equation}
	vanishes at the two Gaussian points and has zero exact mean, while $r(1/2)=-1/12$. For a normal derivative of a quadratic polynomial on a straight face with constant normal, the trace is at most linear. Its zeros at two distinct points then imply a zero derivative along the entire face.
	
	On a regular quadratic curve, the value trace of a physical quadratic can be of degree four in $s$, and the arc-length Jacobian is generally non-polynomial. Two zero value residuals imply a zero quadrature average, while the exact average may differ. The homogeneous normal-derivative condition has a special integral property in two dimensions. For a consistent orientation, $j_s\nn$ is a fixed $90^\circ$ rotation of $\xx'(s)$, hence is linear in $s$. The physical gradient $\nabla P_Q(\xx(s))$ has degree at most two in $s$. The integrand $j_s\nn^{\mathsf T}\nabla P_Q$ therefore has degree at most three, and the two-point Gaussian rule integrates it exactly. In this setting, the two homogeneous derivative constraints imply a zero integrated derivative. The derivative may still vary locally between the sampling points. This argument concerns the normal derivative of the quadratic candidate and does not extend to the nonlinear kinetic flux.
	
	\subsection{Assembly and local solvability}
	\label{sec:assembly}
	
	Use the mean-preserving basis in Eq.~\eqref{eq:poly}. In a quadrilateral cell with one wall face and three fluid-side face neighbors, the neighboring averages give
	\begin{equation}
		D_{j,:}=\frac{1}{|\Omega_j|}\int_{\Omega_j}\bm\phi^{\mathsf T}\dd\Omega,
		\qquad d_{Q,j}=\av Q_j-\av Q_i .
		\label{eq:neighbor}
	\end{equation}
	The target-cell basis is used throughout these integrals. At each Gaussian point let $(\xi_k,\eta_k)$ be its coordinates in that basis, and let $\bm b_k=\bm J^{-1}\nn_k$. The value and normal-derivative rows are
	\begin{align}
		\bm c_{V,k}&=\bm\phi(\xi_k,\eta_k)^{\mathsf T},
		\nonumber\\
		\bm c_{N,k}&=(b_{k,1},b_{k,2},2b_{k,1}\xi_k,
		b_{k,1}\eta_k+b_{k,2}\xi_k,2b_{k,2}\eta_k).
		\label{eq:rows}
	\end{align}
	The derivative row follows from $\nabla_{\xx}P_Q=\bm J^{-\mathsf T}\nabla_{\bm r}P_Q$. The right-hand side of a value row is $Q_{w,k}-\av Q_i$. The homogeneous derivative row has a zero right-hand side.
	
	The three average rows and two wall rows form $\bm C_Q\bm a_Q=\bm e_Q$. For momentum and density, the two wall rows are value rows. For total energy they are derivative rows. The compact procedure also assembles a gradient objective. For each neighbor, its two rows in $\bm A$ are the neighbor averages of $\partial_\xi\bm\phi^{\mathsf T}$ and $\partial_\eta\bm\phi^{\mathsf T}$. The corresponding data in $\bm g_Q$ are the components of $\bm J^{\mathsf T}\av{\nabla_{\xx}Q}_j$. The reconstruction is written as
	\begin{equation}
		\min_{\bm a_Q}\|\bm A\bm a_Q-\bm g_Q\|_2^2
		\quad\text{subject to}\quad
		\bm C_Q\bm a_Q=\bm e_Q .
		\label{eq:cls}
	\end{equation}
	An equivalent multiplier form is
	\begin{equation}
		\begin{bmatrix}
			2\bm A^{\mathsf T}\bm A & \bm C_Q^{\mathsf T}\\
			\bm C_Q & \bm 0
		\end{bmatrix}
		\begin{bmatrix}\bm a_Q\\\bm\nu_Q\end{bmatrix}
		=
		\begin{bmatrix}2\bm A^{\mathsf T}\bm g_Q\\\bm e_Q\end{bmatrix}.
		\label{eq:kkt}
	\end{equation}
	When the five equality rows are independent, they determine the five coefficients uniquely. The gradient objective then has no free direction in which to modify the quadratic candidate. With three independent neighbor rows and one independent wall-average row, one coefficient direction remains available to fit the gradients. Thus the second wall equation increases control at the flux points at the expense of freedom to fit the neighboring gradient data.
	
	The independence of the rows can be demonstrated on a regular stencil. Take the target to be a unit square centered at $(0,0)$ in $(\xi,\eta)$ with the wall at $\eta=-1/2$. Place the three unit-square neighbor centers at $(-1,0)$, $(1,0)$ and $(0,1)$. For homogeneous data, the neighbor averages give $a_1=a_3=0$ and $a_2+a_5=0$. The two wall points have $\xi=\pm1/(2\sqrt3)$. For value rows, subtracting the wall equations gives $a_4=0$, and their mean gives $-a_2/2+a_5/6=0$. For derivative rows, the straight-wall derivative is proportional to $a_2+a_4\xi-a_5$. The difference and mean then give $a_4=0$ and $a_2-a_5=0$. Both systems have only the zero homogeneous solution. The entries of each matrix vary continuously with regular changes in the stencil, so nonsingularity persists for sufficiently small perturbations of this example. For a general curved and stretched stencil, the rank and conditioning depend on the actual moments and normals. If equality rows become redundant, the multiplier system in Eq.~\eqref{eq:kkt} is singular. Consistency of the right-hand side and control of the remaining null space then need separate consideration.
	
	\subsection{Sensitivity and accuracy of the quadratic candidate}
	
	For conditioning analysis, scale each derivative row and its right-hand side by a common positive length. Let $\widehat{\bm C}_Q\bm a_Q=\widehat{\bm e}_Q$ be the resulting system. Perturbations in geometry or data give
	\begin{equation}
		(\widehat{\bm C}_Q+\delta\widehat{\bm C}_Q)\delta\bm a_Q
		=\delta\widehat{\bm e}_Q-\delta\widehat{\bm C}_Q\bm a_Q.
	\end{equation}
	For a nonsingular perturbed matrix, this yields
	\begin{equation}
		\|\delta\bm a_Q\|_2\leq
		\frac{\|\delta\widehat{\bm e}_Q\|_2+
			\|\delta\widehat{\bm C}_Q\|_2\|\bm a_Q\|_2}
		{\sigma_{\min}(\widehat{\bm C}_Q+\delta\widehat{\bm C}_Q)}.
		\label{eq:pos-conditioning}
	\end{equation}
	A small singular value under the chosen scaling allows amplification of geometry or data perturbations. Recovering the physical gradient introduces $\bm J^{-\mathsf T}$ as a separate factor. Both effects can be relevant on stretched meshes. Equation~\eqref{eq:pos-conditioning} identifies a sensitivity mechanism. Its magnitude on a particular grid requires its local matrices and data.
	
	With exact cell averages and geometry moments, compatible wall data and a nonsingular equality matrix, a quadratic field in the reconstruction space is reproduced exactly. Its coefficients satisfy the five equations and are their unique solution. This applies to the density candidate when the density targets are also compatible with the field. Polynomial reproduction concerns the unlimited local reconstruction. The accuracy of the full scheme additionally depends on the pressure-gradient approximation, flux integration, nonlinear selection, and the resolution of the thermal layer. In particular, satisfying the two boundary equations supplies local boundary information but cannot recover fine-scale structure absent from the fluid-side data.
	
	\subsection{Sequential reconstruction and nonlinear selection}
	\label{sec:wall-sequence}
	
	The momentum and total-energy candidates are constructed first. Write $\widetilde P_Q$ for the polynomial retained after nonlinear selection. The two density targets use the selected energy and prescribed temperature,
	\begin{equation}
		\rho_{w,k}^{*}=\frac{(\gamma-1)\widetilde P_{\mathcal E}(\xx_{G,k})}
		{R_gT_w},
		\qquad P_{\rho}^{(2)}(\xx_{G,k})=\rho_{w,k}^{*}.
		\label{eq:density}
	\end{equation}
	The density candidate then undergoes its own selection. The implementation first defines, for every face $f$ of cell $i$,
	\begin{equation}
		A_f=\frac{(p_L+p_R)|p_L-p_R|}{p_Lp_R}
		+(M_{n,L}-M_{n,R})^2+
		\|\bm M_{t,L}-\bm M_{t,R}\|_2^2,
		\qquad
		\alpha_i=\prod_{f\subset\partial\Omega_i}\frac{1}{1+A_f^2}.
		\label{eq:feedback}
	\end{equation}
	Here $M_n$ and $\bm M_t$ are the normal and tangential velocity components divided by the local sound speed, and the two states are those supplied to the face calculation. The product includes faces adjacent to auxiliary wall states. In the branch $\alpha_i<0.9$, the nonlinear selection replaces the quadratic candidate by the modified linear Green--Gauss representation
	\begin{equation}
		P_Q^{\rm use}(\xx)=\av Q_i+
		\alpha_i\bm G_Q^{\rm GG}\cdot(\xx-\xx_i),
		\label{eq:feedback-fallback}
	\end{equation}
	with all quadratic coefficients set to zero. Otherwise the quadratic candidate is retained. Thus the two wall constraints are imposed on the quadratic candidate, but they need not be satisfied by the polynomial used by the numerical flux after nonlinear selection.
	
	For a positive selected density, the reconstructed temperature at $\xx_{G,k}$ is $T_w$ when the selected momenta are zero and the density retains the target in Eq.~\eqref{eq:density}. If nonlinear selection changes the momentum or density trace, equality with $T_w$ is no longer ensured, and selection can also change the normal derivative of total energy. The two equality rows provide no positivity guarantee between the samples. The saved terminal fields do not contain the face-by-face values of $\alpha_i$, the branch selected, or a wall-point admissibility log. Consequently, the present results establish the response of the complete POS scheme, but do not quantify how often the constrained quadratic reaches the wall flux.

	\subsection{Coupling to the kinetic wall flux}
	
	Wall reconstruction and kinetic mass balance play different roles. With $u_n=\vv\cdot\nn_k>0$ denoting particles traveling from the fluid toward the wall, the reconstructed inner state gives the incident distribution
	\begin{equation}
		f^I(t)=g^I[1-\tau_{\rm phy}(\bm a^I\cdot\vv+A^I)
		-t\bm a^I\cdot\vv] .
	\end{equation}
	The outgoing half-space $u_n<0$ is assigned a stationary Maxwellian at $T_w$, with a scalar time coefficient,
	\begin{equation}
		f^R(t)=\rho_R\widehat g_w(1+A_Rt),\qquad
		\lambda_w=(2R_gT_w)^{-1}.\label{eq:reflected}
	\end{equation}
	Here $\widehat g_w$ has unit density and $\bm a^I,A^I$ have the meaning defined in \eqnref{eq:ce}. At fixed wall velocity and temperature, only the Maxwellian density changes in time, giving the scalar $A_R=(\partial_t\rho_R)/\rho_R$. Following \cite{zhang2026}, define the incident mass coefficients
	\begin{align}
		m_0&=\int_{u_n>0}u_nf^I(0)\dd\Xi,\nonumber\\
		m_1&=\int_{u_n>0}u_n\partial_tf^I\dd\Xi.
		\label{eq:incidentmass}
	\end{align}
	Since $\int_{u_n<0}u_n\widehat g_w\dd\Xi=-\sqrt{R_gT_w/(2\pi)}$, mass balance gives
	\begin{equation}
		\rho_R=m_0\sqrt{\frac{2\pi}{R_gT_w}},\qquad
		A_R=\frac{m_1}{m_0}.
		\label{eq:massbalance}
	\end{equation}
	These expressions require $m_0>0$ for an admissible outgoing density. Because both half-distributions are linear in time, the constant and linear mass-flux coefficients cancel separately. This algebraic cancellation does not establish positivity of the approximate distributions. The wall flux is
	\begin{equation}
		\bm F_w(t)=\int_{u_n>0}u_nf^I\bm\psi\dd\Xi
		+\int_{u_n<0}u_nf^R\bm\psi\dd\Xi .\label{eq:wallflux}
	\end{equation}
	In the present wall solver, the energy-flux component and its time coefficient are divided by the prescribed Prandtl number. The outgoing density $\rho_R$ is determined by kinetic mass balance, while $\rho_{w,k}^{*}$ supplies the target for the fluid-side reconstruction.
	
	The auxiliary wall state used for gradient evolution is a further, separate construction. The current routine initializes it as $(\rho_R,0,0,\rho_RC_vT_w)^{\mathsf T}$ and supplies the time coefficient $(\rho_RA_R,0,0,0)^{\mathsf T}$. In this auxiliary update the energy component is held fixed over the step. The flux itself uses the half-space energy moments in Eq.~\eqref{eq:wallflux}.

	\section{Cylinder configuration and meshes}
	\label{sec:case}
	
	\subsection{Flow conditions}
	
	The test is laminar flow over the upstream half of a stationary circular cylinder. The cylinder is centered at the origin, the free stream is directed along positive $x$, and the computational diameter is $D=1\,\mathrm{m}$. The domain is bounded by the upstream cylinder wall, a curved exterior boundary and two straight truncation boundaries at $x=0$ outside the cylinder. The mesh extends over $-1\leq x/D\leq 0$ and $-2\leq y/D\leq 2$. A nonreflecting condition is imposed on the curved exterior boundary and symmetry conditions are imposed on the straight truncations. The forward stagnation point is at $(x/D,y/D)=(-1/2,0)$. The signed surface angle $\theta$ is zero there. In the output convention, positive angles lie on the $y<0$ side.
	
	Table~\ref{tab:conditions} gives the parameters of the cylinder calculation. The same computational diameter and free-stream parameters are used for the three sets of results. The gas is calorically perfect, with
	\begin{equation}
		U_\infty=M_\infty\sqrt{\gamma R_gT_\infty},\qquad
		\rho_\infty=\frac{Re_D\,\mu(T_\infty)}{U_\infty D},
		\qquad p_\infty=\rho_\infty R_gT_\infty .
	\end{equation}
	The Reynolds number based on radius is $Re_D/2=1.835\times10^5$. The temperature dependence of viscosity is
	\begin{equation}
		\mu(T)=\mu_s\left(\frac{T}{T_s}\right)^{3/2}
		\frac{T_s+S}{T+S}.
	\end{equation}
	
	\begin{table}[!tb]
		\centering
		\caption{Cylinder flow conditions and gas properties.}
		\label{tab:conditions}
		\begin{tabular}{llrl}
			\toprule
			Quantity & Symbol & Value & Unit \\
			\midrule
			Computational diameter & $D$ & 1.0 & $\mathrm{m}$ \\
			Free-stream Mach number & $M_\infty$ & 8.03 & -- \\
			Diameter-based Reynolds number & $Re_D$ & $3.67\times10^5$ & -- \\
			Free-stream temperature & $T_\infty$ & 124.94 & $\mathrm{K}$ \\
			Wall temperature & $T_w$ & 294.44 & $\mathrm{K}$ \\
			Specific-heat ratio & $\gamma$ & 1.4 & -- \\
			Specific gas constant & $R_g$ & 287.15 & $\mathrm{J\,kg^{-1}\,K^{-1}}$ \\
			Prandtl number & $\mathrm{Pr}$ & 0.72 & -- \\
			Sutherland reference viscosity & $\mu_s$ & $1.785\times10^{-5}$ & $\mathrm{Pa\,s}$ \\
			Sutherland reference temperature & $T_s$ & 288.15 & $\mathrm{K}$ \\
			Sutherland constant & $S$ & 110.4 & $\mathrm{K}$ \\
			\bottomrule
		\end{tabular}
	\end{table}
	
	\subsection{Controlled variation of the wall-normal mesh}
	
	The UR, LLU, and LLP mesh families adapt the near-wall node-placement classifications used by Carlson, Vatsa, and White for wall-function grids \cite{carlson2015}.  In that work, uniform refinement provides a reference sequence, log-layer unpacked (LLU) spacing increases the first off-wall distance while retaining the grid dimensions, and log-layer packed (LLP) spacing removes near-wall layers while retaining the outer-node distribution.  The present calculation uses these names only to distinguish the geometric redistribution of wall-normal nodes.  It does not use a wall function or the turbulent wall coordinate $y^+$, and it does not transfer the RANS wall-function conclusions of Carlson et al. to the laminar hypersonic cylinder case.
	
\begin{figure}[!tb]
		\centering
		\panel{.31}{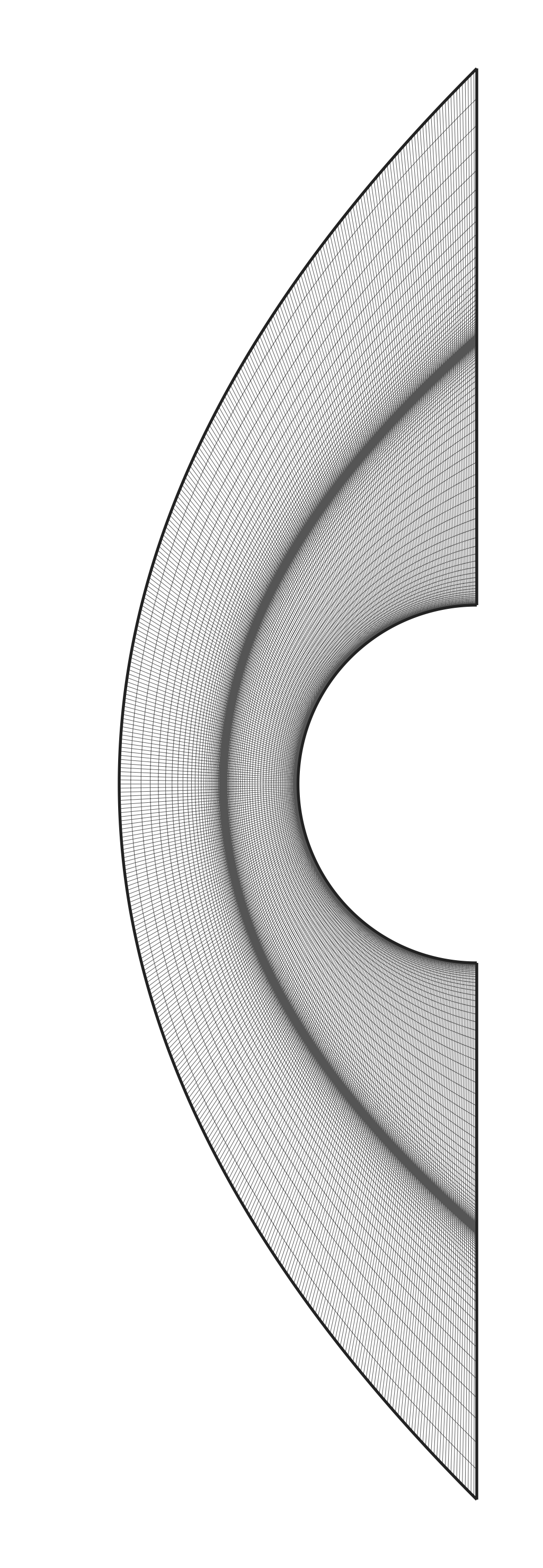}{Straight-sided reference mesh}
		\hspace{.08\linewidth}
		\panel{.31}{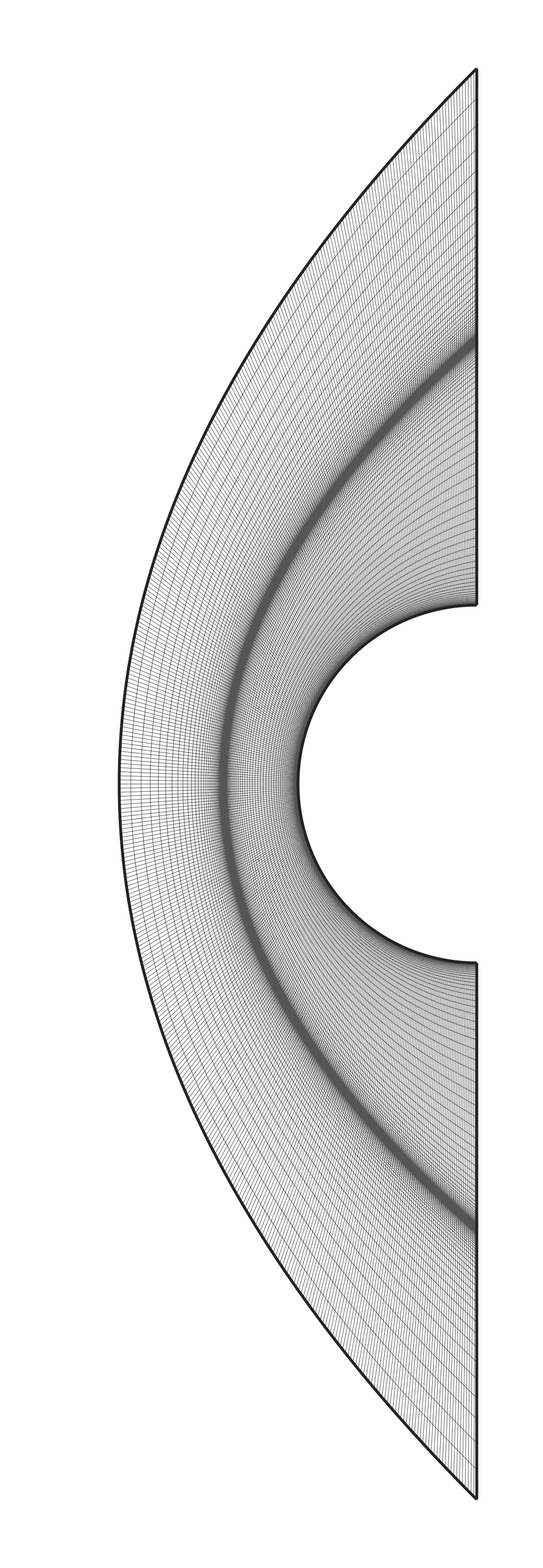}{Quadratic reference mesh}
		\caption{Reference mesh for the upstream half-cylinder in the two geometry representations.}
		\label{fig:meshes}
	\end{figure}

	The seven mesh designs share 400 circumferential intervals and the same outer mesh within each geometry representation. GG and GC use straight-sided quadrilaterals. POS uses the corresponding quadratic geometry with the same corner-node layout. Figure~\ref{fig:meshes} shows both representations for the reference design. The geometry distinction across schemes is retained throughout the comparisons.
	
	% Figure 2
	
	The modifications are confined to the first wall-normal band, of thickness $0.005D$ along the forward stagnation line. The reference has 32 intervals across this band. UR reduces the interval count throughout the band. LLU increases the first-cell height by redistributing the spacing while retaining 32 intervals. LLP joins the first two or four reference intervals and retains the reference nodes beyond the merged region. This distinction is important because two meshes can have nearly equal first-cell heights while resolving the rest of the band quite differently.
	
	The subscripts 2 and 4 denote the two design levels. The realized first heights are about $2.13$--$2.14$ and $4.93$--$4.95$ times the reference height. These labels describe the mesh modifications. They do not specify a global refinement ratio. The three families therefore probe sensitivity to where resolution is removed, and a global order of convergence cannot be inferred from their level indices. Table~\ref{tab:mesh-designs} gives the measured stagnation-line first height $h_1$, number of intervals $N_b$ in the modified band, and total number of control volumes $N_c$.
	
	\begin{table}[!tb]
		\centering
		\caption{Near-wall mesh designs. Heights are measured along the forward stagnation line.}
		\label{tab:mesh-designs}
		\begin{tabular}{lrrrr}
			\toprule
			Mesh & $N_b$ & $10^{5}h_1/D$ & $h_1/h_{1,\mathrm{ref}}$ & $N_c$ \\
			\midrule
			Reference & 32 & 1.0013 & 1.000 & 50,800 \\
			$\mathrm{UR}_2$ & 16 & 2.1461 & 2.143 & 44,400 \\
			$\mathrm{LLU}_2$ & 32 & 2.1329 & 2.130 & 50,800 \\
			$\mathrm{LLP}_2$ & 31 & 2.1461 & 2.143 & 50,400 \\
			$\mathrm{UR}_4$ & 8 & 4.9522 & 4.946 & 41,200 \\
			$\mathrm{LLU}_4$ & 32 & 4.9366 & 4.930 & 50,800 \\
			$\mathrm{LLP}_4$ & 29 & 4.9512 & 4.945 & 49,600 \\
			\bottomrule
		\end{tabular}
	\end{table}
	
	% Figure 3
	\begin{figure}[!tb]
		\centering
		\panel{.31}{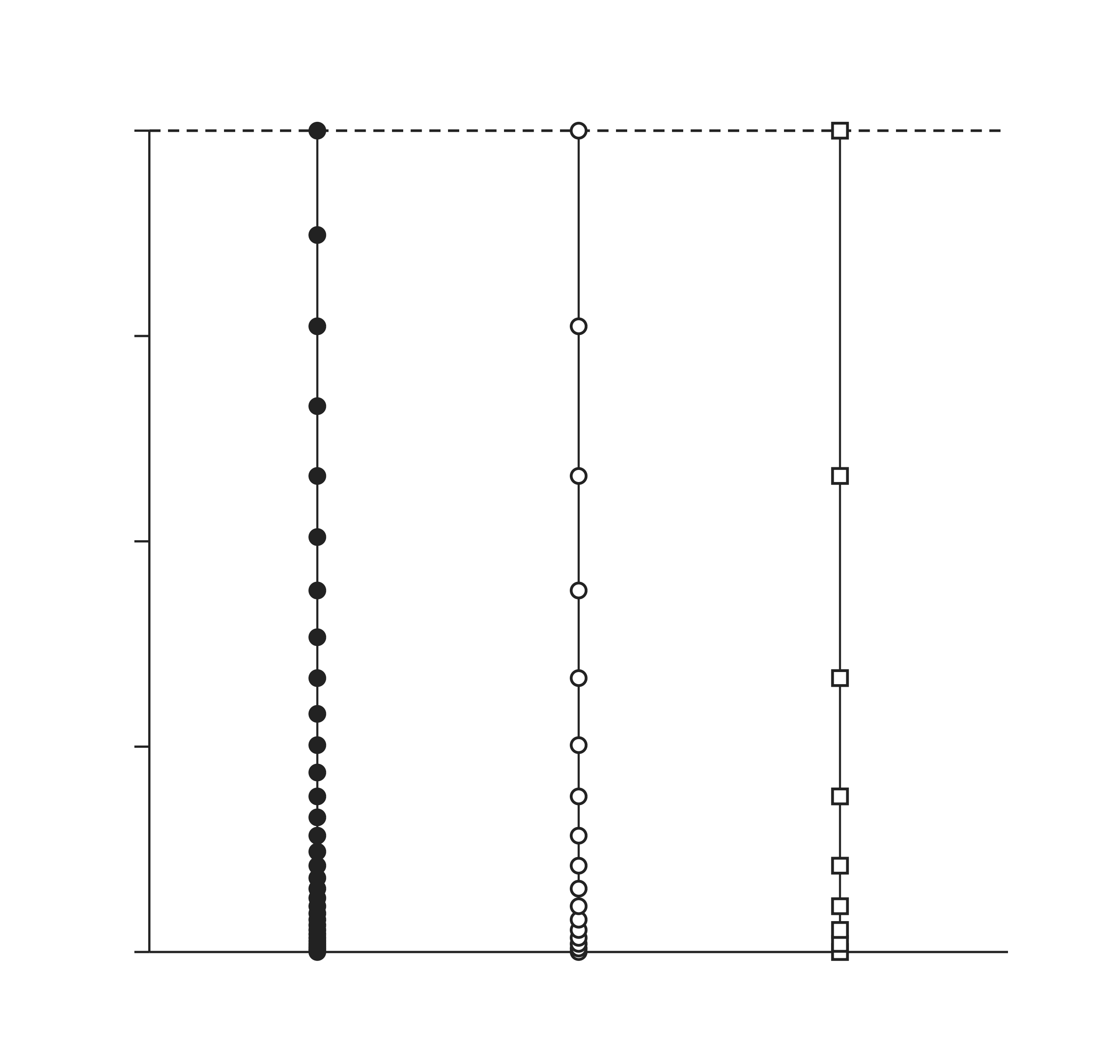}{UR, full band}\hfill
		\panel{.31}{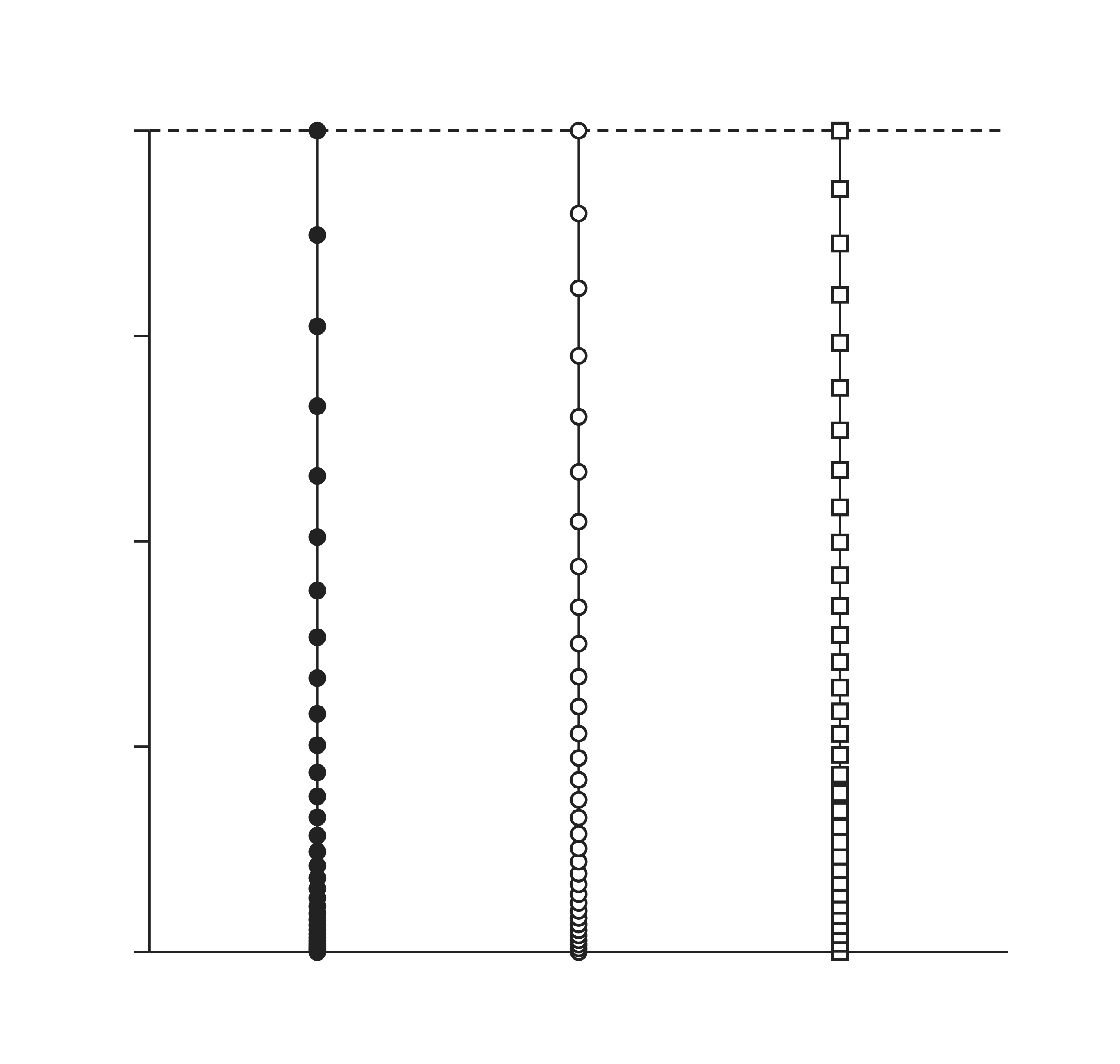}{LLU, full band}\hfill
		\panel{.31}{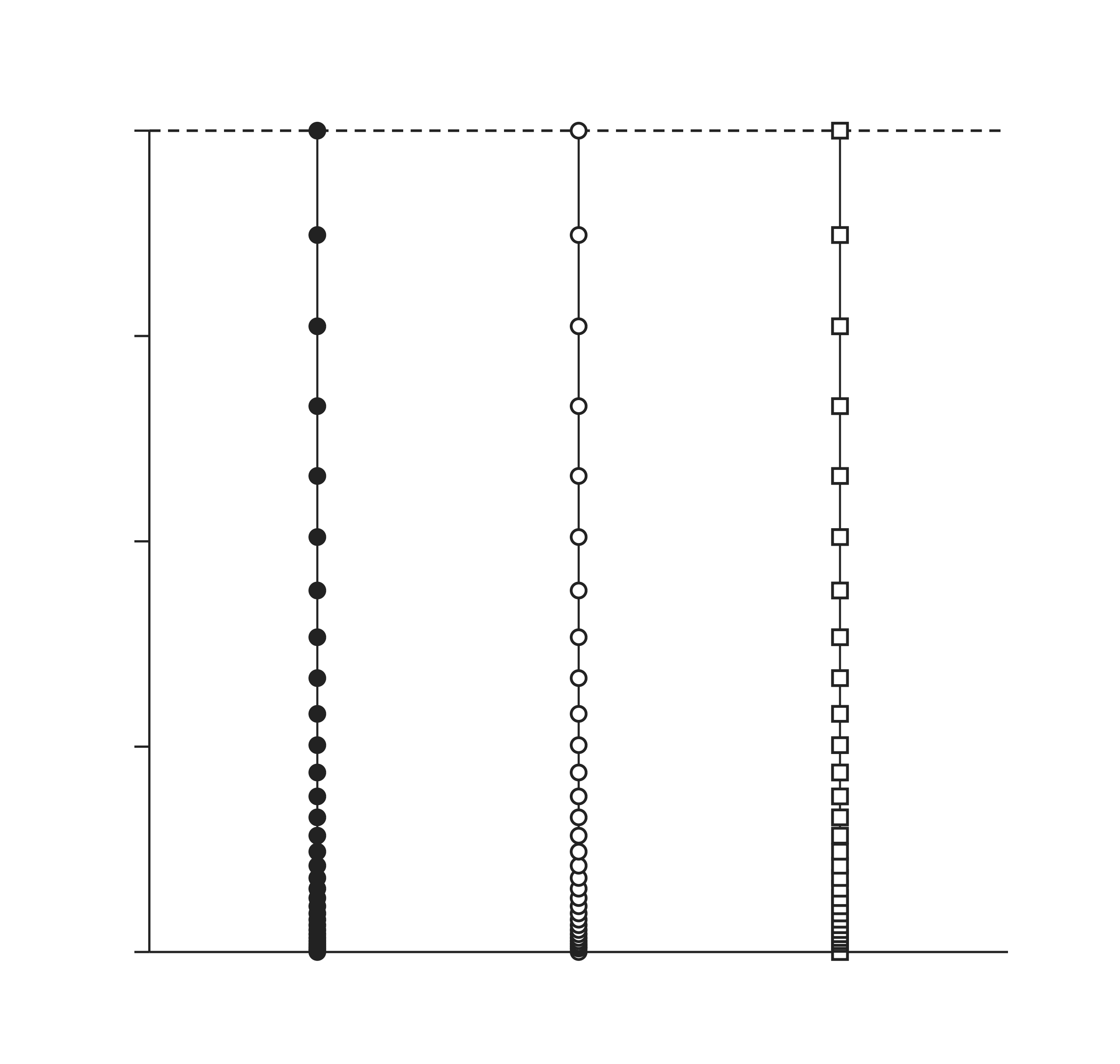}{LLP, full band}\par\medskip
		\panel{.31}{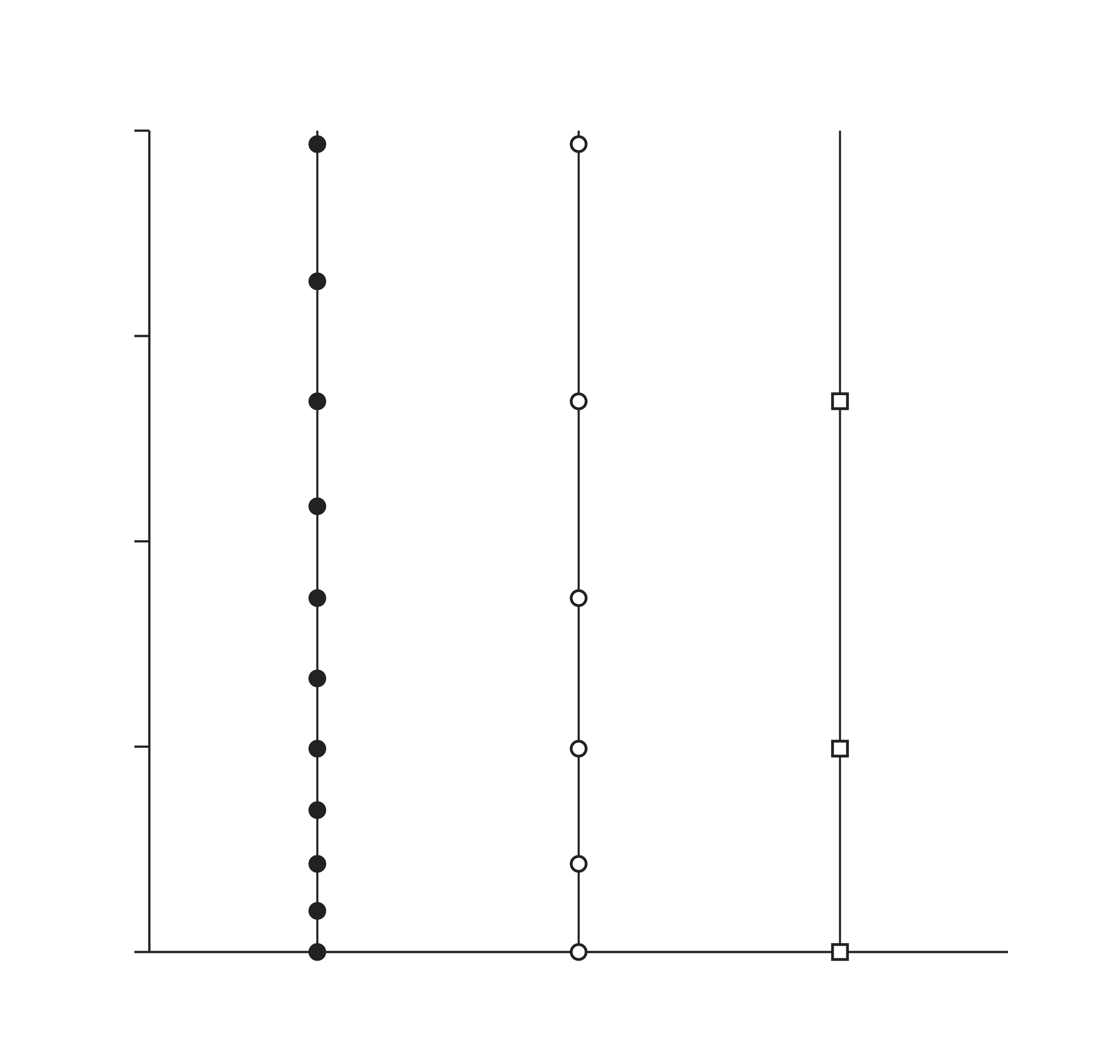}{UR, near-wall detail}\hfill
		\panel{.31}{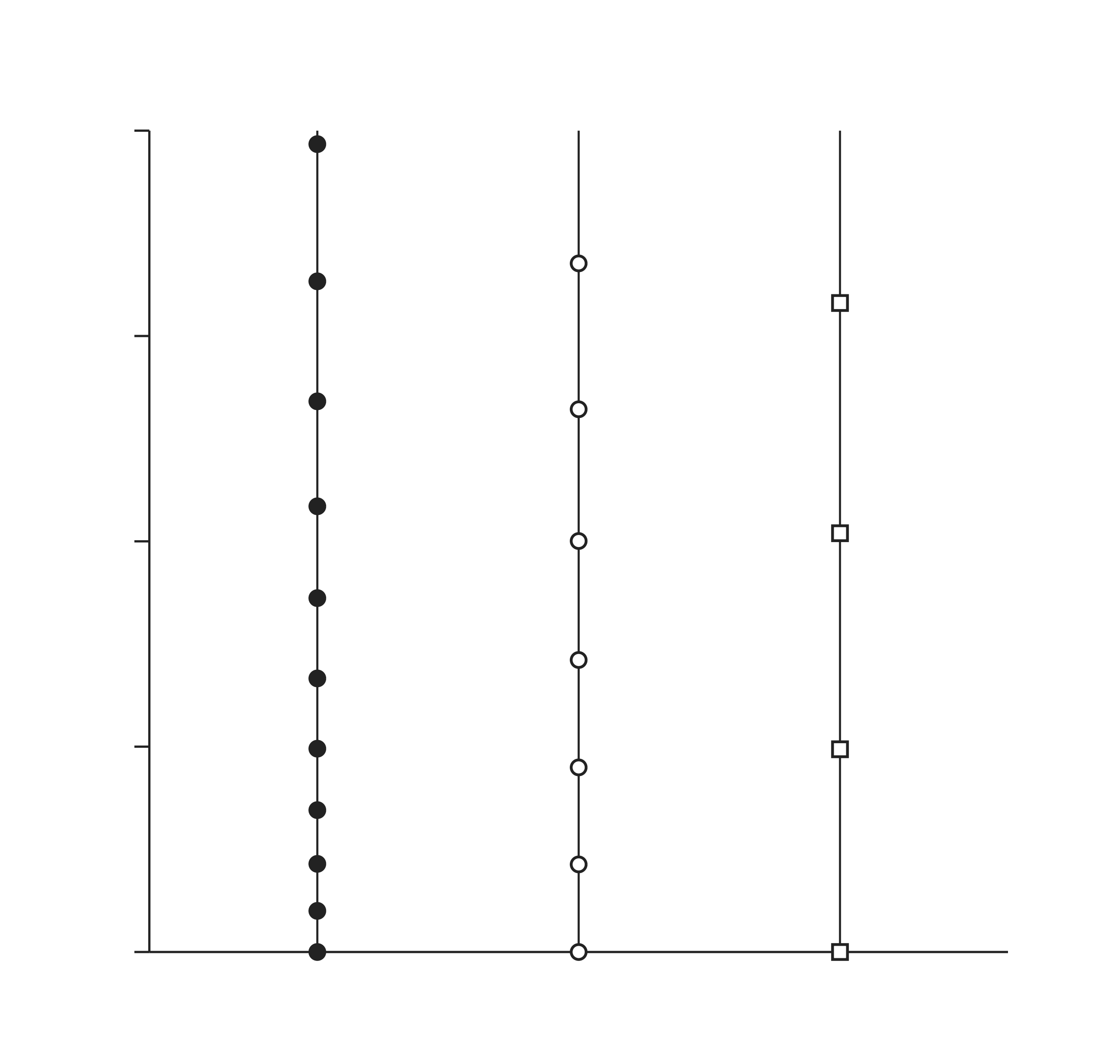}{LLU, near-wall detail}\hfill
		\panel{.31}{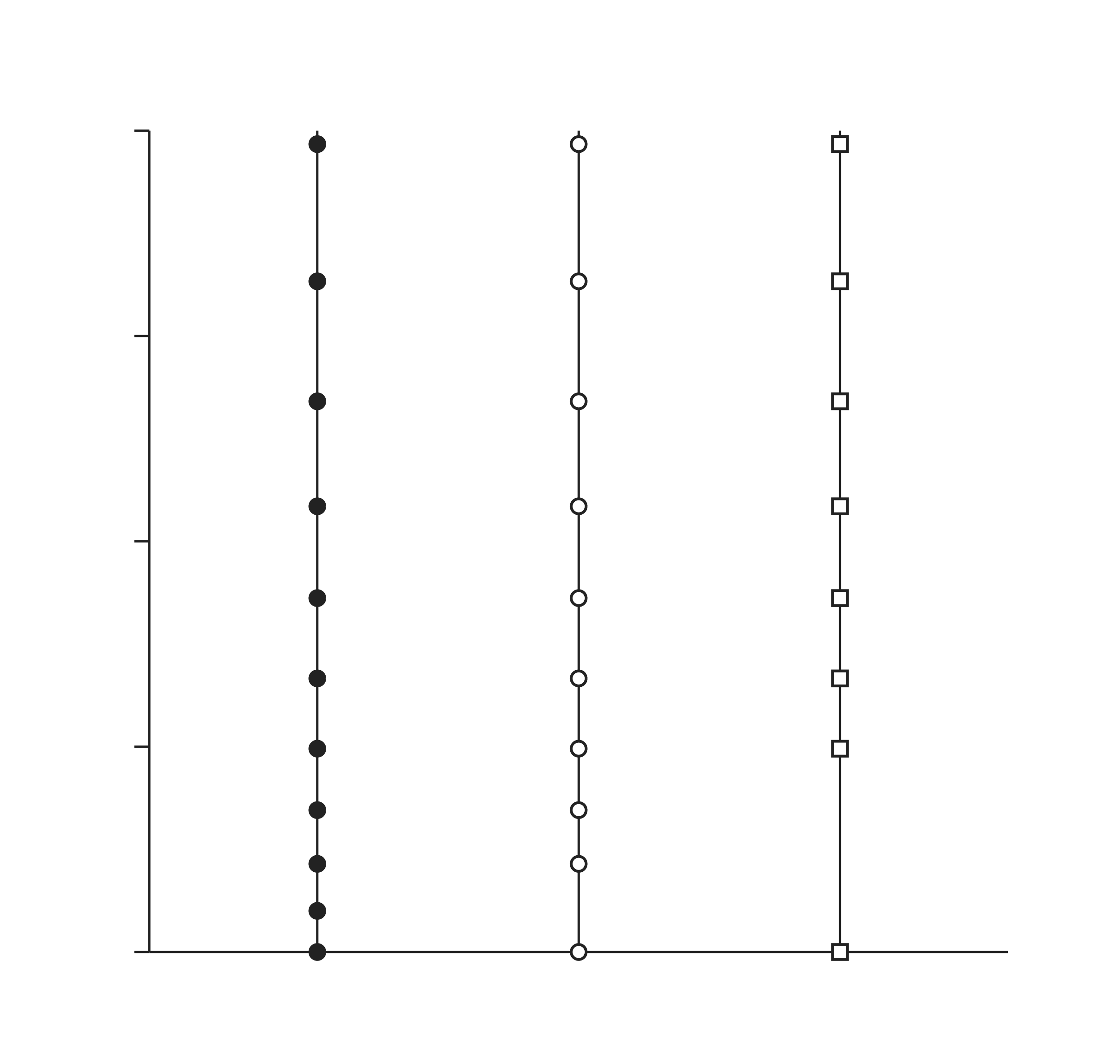}{LLP, near-wall detail}
		\caption{Wall-normal nodes along the forward stagnation line. UR reduces the interval count throughout the modified band; LLU redistributes nodes while retaining the interval count; LLP merges only the innermost reference intervals. The top row covers the modified band and the bottom row enlarges its inner part. The abscissa separates the reference and the two design levels. The ordinate is wall-normal distance divided by the modified-band thickness.}
		\label{fig:nodes}
	\end{figure}
	
	Figure~\ref{fig:nodes} shows why $h_1$ and $N_b$ convey different information. At level 4, UR and LLP have nearly the same first height, but there are eight intervals in UR and 29 in LLP. LLU retains the reference number of control volumes and changes their wall-normal allocation. Thus this test also separates a change in resolution from a reduction in the total number of control volumes.
	
	\subsection{Terminal data and scope of the numerical evidence}
	
	All figures use the selected terminal surface and flow files archived for the 21 scheme--mesh calculations; no temporal averaging is applied. The archive records the final field and wall-output files used in postprocessing, but it does not preserve complete residual histories, stagnation heat-flux histories, a time-step study, or mesh refinements beyond the reference design. The results therefore document the response of the selected terminal calculations to the stated mesh modifications. They do not establish a residual-based steady stopping criterion, time-step independence, circumferential-grid independence, or asymptotic spatial convergence.
	
	\section{Results and discussion}
	\label{sec:results}
	
	\subsection{Surface quantities and comparison measures}
	
	The dimensionless outputs used for the surface comparisons are
	\begin{equation}
		p^{*}=\frac{p_w}{0.9209\rho_\infty U_\infty^2},
		\qquad
		q^{*}=\frac{q_w}{0.003655\rho_\infty U_\infty^3}.
		\label{eq:surface-normalization}
	\end{equation}
	Here $q_w$ denotes the numerical wall energy flux per unit wall-face length, positive toward the wall in the reported convention. The constants $0.9209$ and $0.003655$ are adopted unchanged from the cylinder benchmark normalization used by Zhang et al.\ \cite{zhang2026}; they are not fitted to the present calculations. The same normalizations apply to all three schemes. The pressure observable $p_w$ follows each implementation's surface output. GG computes pressure from the stored interface state. GC computes it from the wall-adjacent cell-average conservative state. POS averages the fluid-side pressure evaluated at the two wall Gaussian points, using their arc-length quadrature weights. The pressure curves therefore compare reported surface diagnostics that use different state samples. Each surface file gives one record per wall face, at angular locations from $-89.775^\circ$ to $89.775^\circ$ with $0.45^\circ$ spacing.
	
	For a surface quantity $z=p^{*}$ or $q^{*}$, define the near-stagnation value $\bar z_s=[z(-0.225^\circ)+z(0.225^\circ)]/2$. The signed change and relative curve difference for a mesh $m$ are
	\begin{align}
		\delta_s(z;m)&=100\frac{\bar z_{s,m}-\bar z_{s,\mathrm{ref}}}
		{\bar z_{s,\mathrm{ref}}},\\
		E_{\mathcal I}(z;m)
		&=100\frac{\left[\sum_{j\in\mathcal I}
			(z_{m,j}-z_{\mathrm{ref},j})^2\right]^{1/2}}
		{\left[\sum_{j\in\mathcal I}z_{\mathrm{ref},j}^2\right]^{1/2}}.
		\label{eq:mesh-metrics}
	\end{align}
	The reference is taken from the same scheme as $m$. The index set $\mathcal I$ includes either all 400 faces or the 66 faces satisfying $|\theta|\leq15^\circ$. The signed change indicates the direction of the response close to stagnation, and the curve difference measures its extent over an angular range. Both are mesh-sensitivity measures relative to a scheme-specific numerical reference; they are neither experimental errors nor differences between schemes.
	
	For the comparison with the digitized experiment, a computed curve is interpolated linearly to the experimental angles. If $z_j$ and $z_j^{\rm exp}$ denote the computed and experimental values at $N$ sampled angles, the reported signed bias and RMS difference are
	\begin{equation}
		B(z)=\frac{1}{N}\sum_{j=1}^{N}(z_j-z_j^{\rm exp}),
		\qquad
		R(z)=\left[\frac{1}{N}\sum_{j=1}^{N}(z_j-z_j^{\rm exp})^2\right]^{1/2}.
		\label{eq:experiment-metrics}
	\end{equation}
	Each digitized point receives equal weight. These are differences at the supplied experimental locations, not uncertainty-weighted estimates of physical error. The experimental angles and coverage differ between pressure and heat flux.
	
	\subsection{Reference meshes and experimental comparison}
	
\begin{figure}[!tb]
		\centering
		\panel{.48}{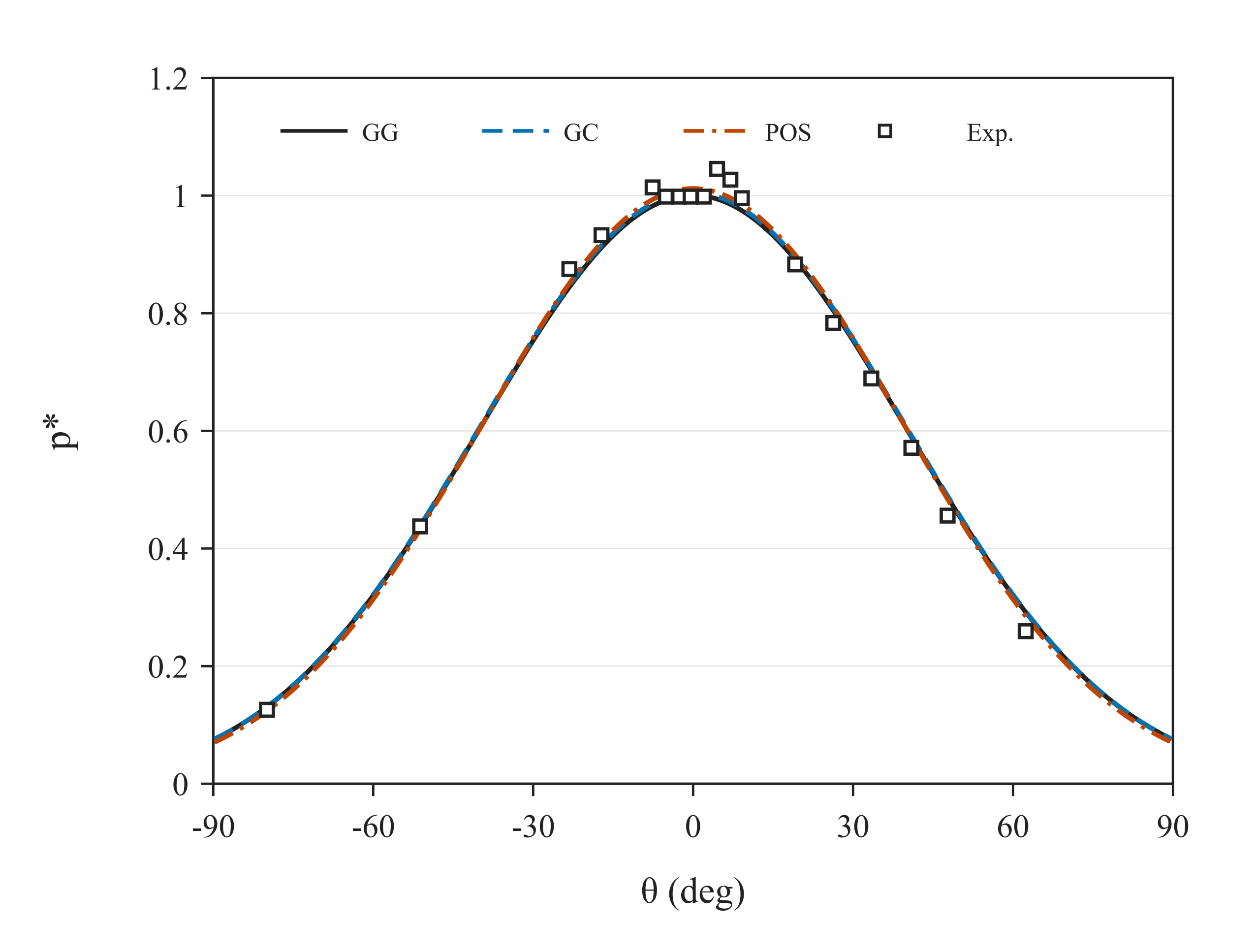}{Wall pressure}\hfill
		\panel{.48}{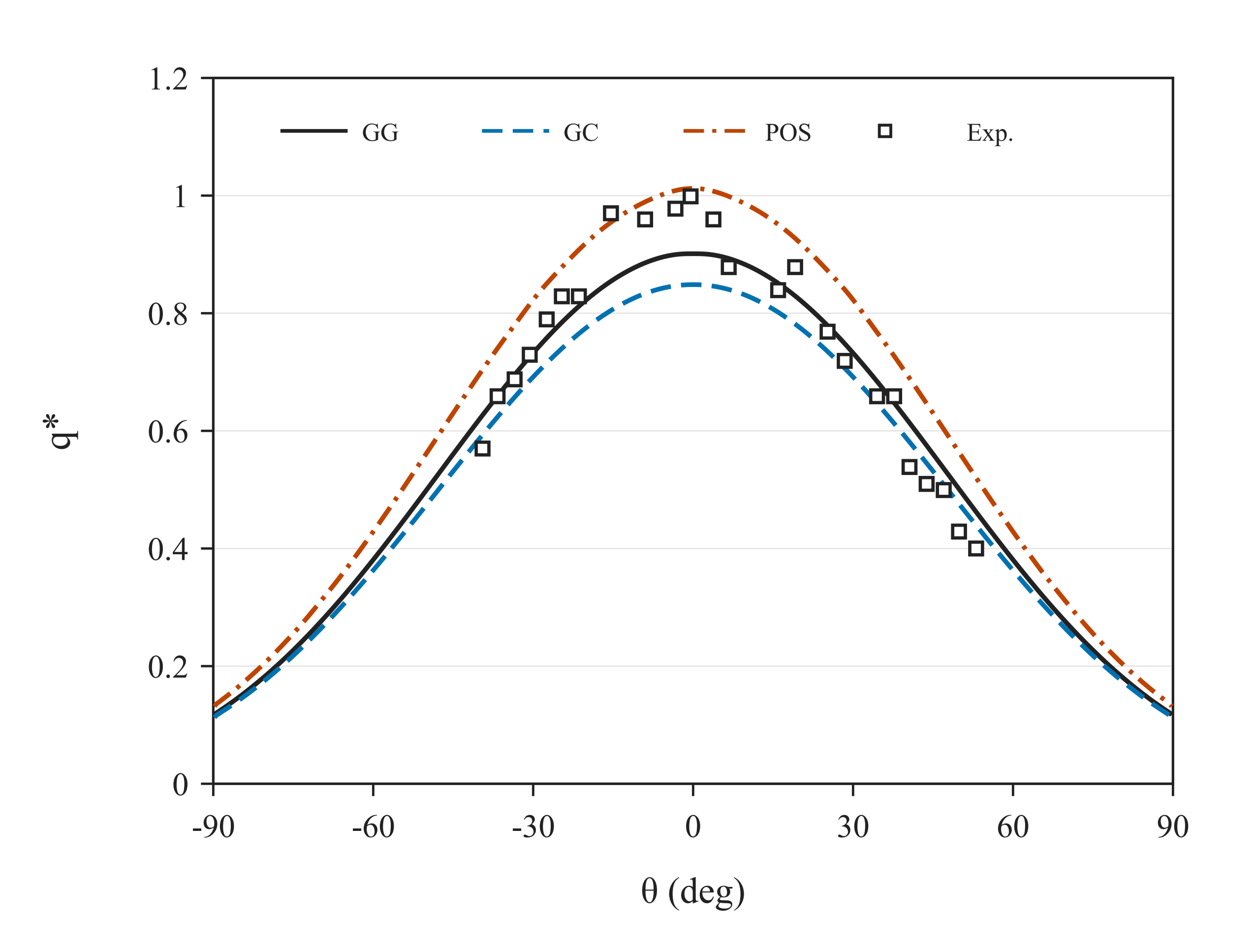}{Wall heat flux}
		\caption{Reference-mesh surface distributions and digitized experimental points. GG is shown by the solid black line, GC by the dashed blue line and POS by the dash-dotted orange line. Open squares denote the digitized experimental data reported by Wieting and Holden \cite{wieting1987} and used in the cylinder benchmark of Zhang et al.\ \cite{zhang2026}.}
		\label{fig:experiment}
	\end{figure}

	Figure~\ref{fig:experiment} compares the three numerical reference curves with the experimental series used for the cylinder benchmark in Zhang et al.\ \cite{zhang2026,wieting1987}. The supplied digitized coordinates contain 18 pressure points and 24 heat-flux points; pressure spans a wider angular range. The archive does not contain the original figure or table from which the coordinates were digitized, nor a digitization or measurement uncertainty. Accordingly, the bias and RMS values below are descriptive comparisons to the supplied points and do not support a statistical accuracy ranking.
	
	% Figure 4
	
	\begin{table}[!tb]
		\centering
		\caption{Reference-mesh quantities. The left group gives near-stagnation predictions. The right group gives differences at the digitized experimental locations: 18 pressure points and 24 heat-flux points. Bias is the mean computed value minus the experimental value; RMS is the root-mean-square difference.}
		\label{tab:benchmark}
		\begin{tabular}{lrrrrrr}
			\toprule
			& \multicolumn{2}{c}{Near-stagnation predictions} & \multicolumn{4}{c}{Differences at sampled experimental locations} \\
			\cmidrule(lr){2-3}\cmidrule(lr){4-7}
			Scheme & $\bar p_s^{*}$ & $\bar q_s^{*}$ & $p^{*}$ bias & $p^{*}$ RMS & $q^{*}$ bias & $q^{*}$ RMS \\
			\midrule
			GG & 1.0004 & 0.9012 & -0.0039 & 0.0240 & -0.0048 & 0.0538 \\
			GC & 1.0045 & 0.8486 & -0.0006 & 0.0231 & -0.0459 & 0.0765 \\
			POS & 1.0120 & 1.0122 & 0.0016 & 0.0203 & 0.0842 & 0.0944 \\
			\bottomrule
		\end{tabular}
	\end{table}
	
	The three reported pressure diagnostics give near-stagnation values between 1.0004 and 1.0120, whereas the corresponding heat-flux predictions range from 0.8486 to 1.0122. At the supplied heat-flux points, GC has a negative mean difference and POS a positive mean difference; GG has the smallest RMS difference among the three reference curves. The numerical ordering of the pressure bias and RMS entries in Table~\ref{tab:benchmark} is reported for completeness only. Because GG, GC and POS sample different numerical states for that output, it is not an interscheme pressure-accuracy ranking.
	
	\subsection{Relation between mesh sensitivity and differences at experimental locations}
	
\begin{figure}[!tb]
		\centering
		\includegraphics[width=.76\linewidth]{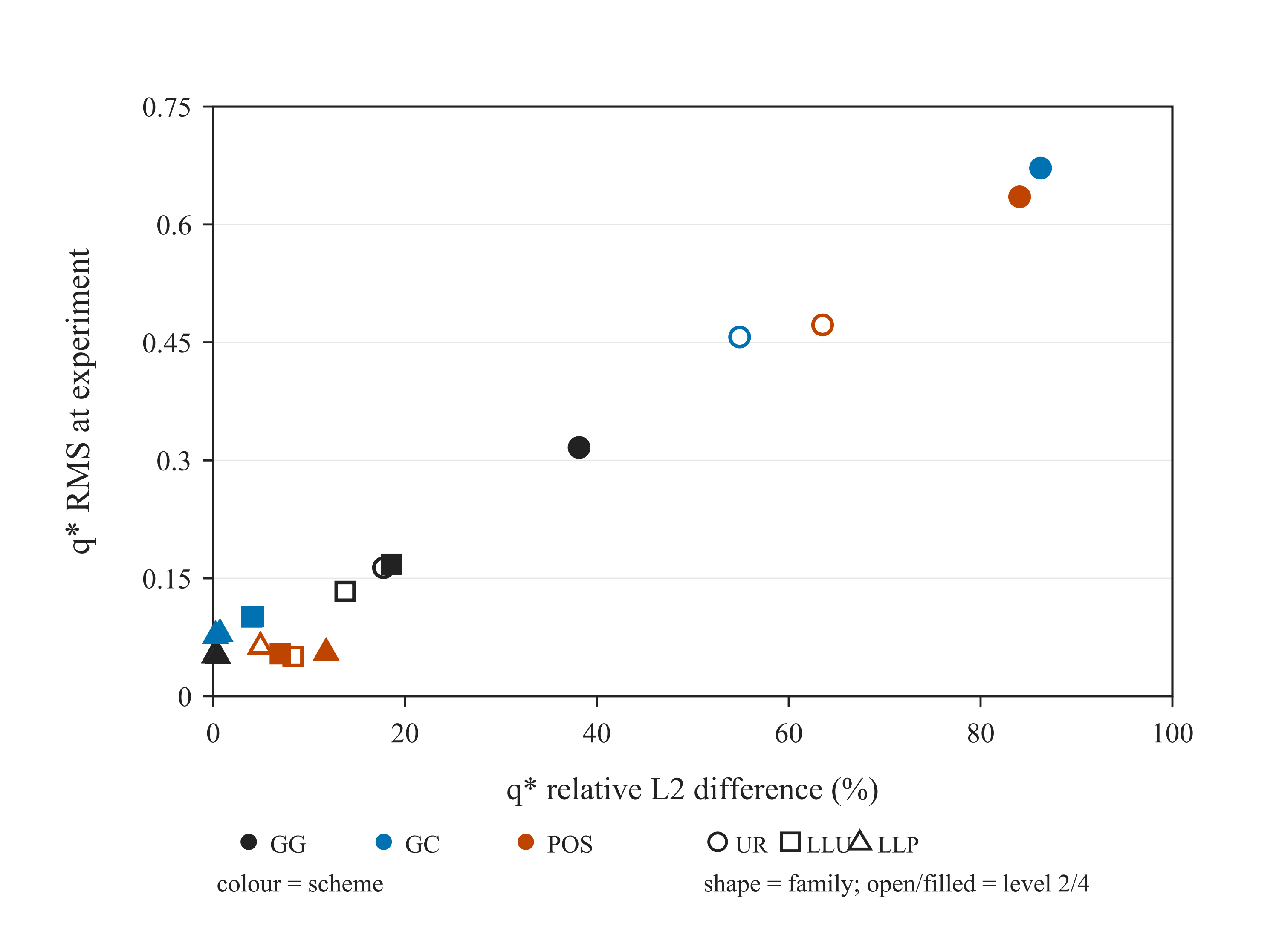}
		\caption{Heat-flux mesh sensitivity and difference at the digitized experimental locations. Each marker denotes one modified scheme--mesh calculation. The horizontal coordinate is the full-curve relative $L_2$ difference from the scheme-specific reference; the vertical coordinate is the RMS difference at the 24 digitized heat-flux angles. Colour identifies the complete scheme, shape identifies the mesh family, and open/filled markers identify levels 2/4.}
		\label{fig:sensitivity-experiment}
	\end{figure}

	Figure~\ref{fig:sensitivity-experiment} places each of the 18 modified scheme--mesh calculations by two distinct quantities: the full-curve heat-flux difference from its own numerical reference, $E_{\rm all}(q^{*};m)$, and the heat-flux RMS difference at the digitized experimental angles, $R(q^{*})$. The abscissa measures response to a mesh modification within a scheme; the ordinate measures agreement with the supplied experimental points. Neither quantity is an interscheme difference.
	
	% Figure 5
	
	The figure prevents the two comparisons from being conflated. For example, POS--LLU2 differs from the POS reference by 8.33\% in the full-curve metric, but its RMS difference at the supplied heat-flux locations is 0.0508, below the POS-reference value of 0.0944. Thus a modified calculation can depart from its numerical reference and nevertheless lie closer to the supplied points. Conversely, the large UR responses coincide with substantially larger RMS differences for the present data. These observations describe this digitized-data comparison only: without experimental and digitization uncertainties, they are not a statistical validation or a general accuracy ranking.
	
	\subsection{Heat-flux sensitivity to the location of mesh modification}
	
\begin{figure}[!tb]
		\centering
		\panel{.32}{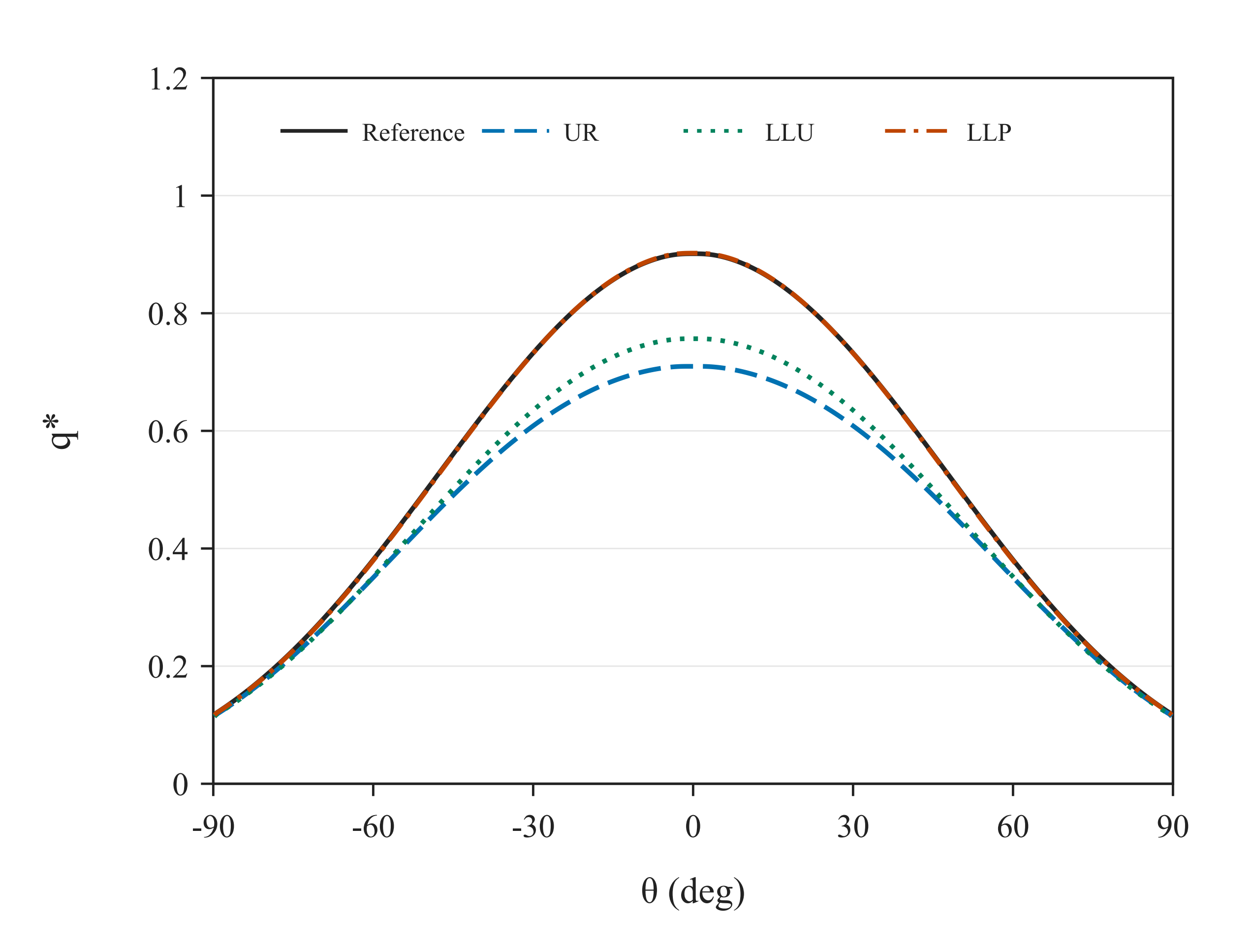}{GG, level 2}\hfill
		\panel{.32}{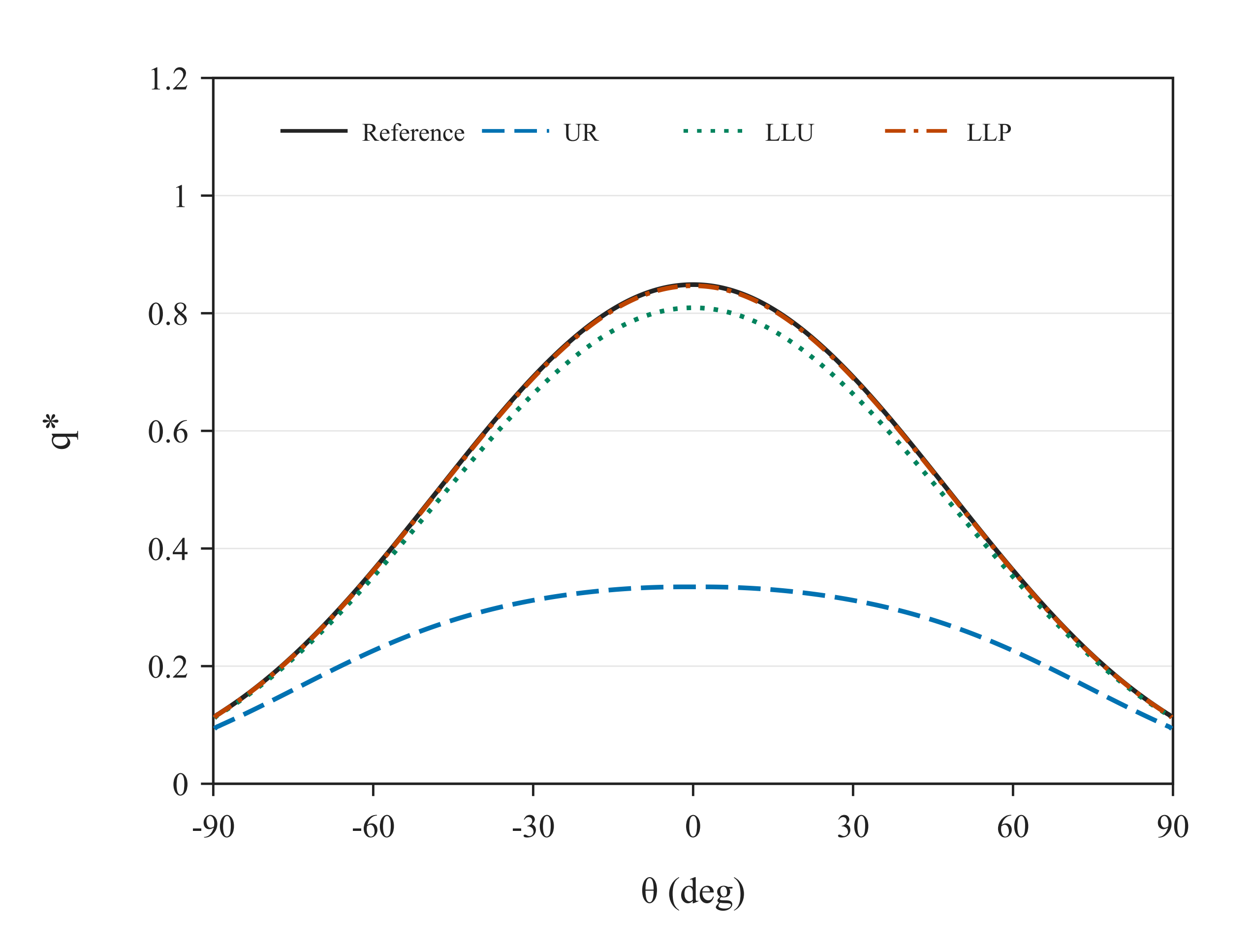}{GC, level 2}\hfill
		\panel{.32}{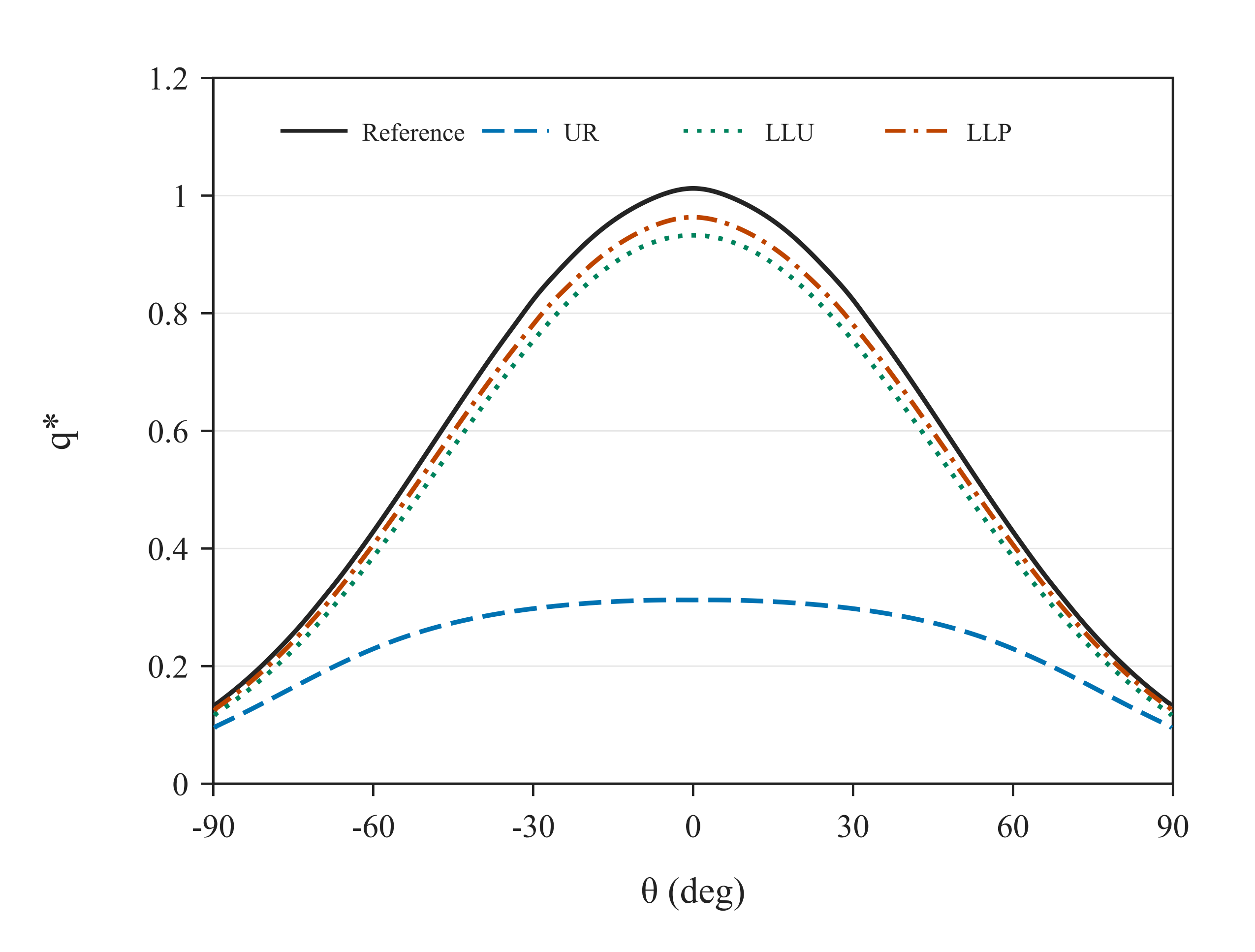}{POS, level 2}\par\medskip
		\panel{.32}{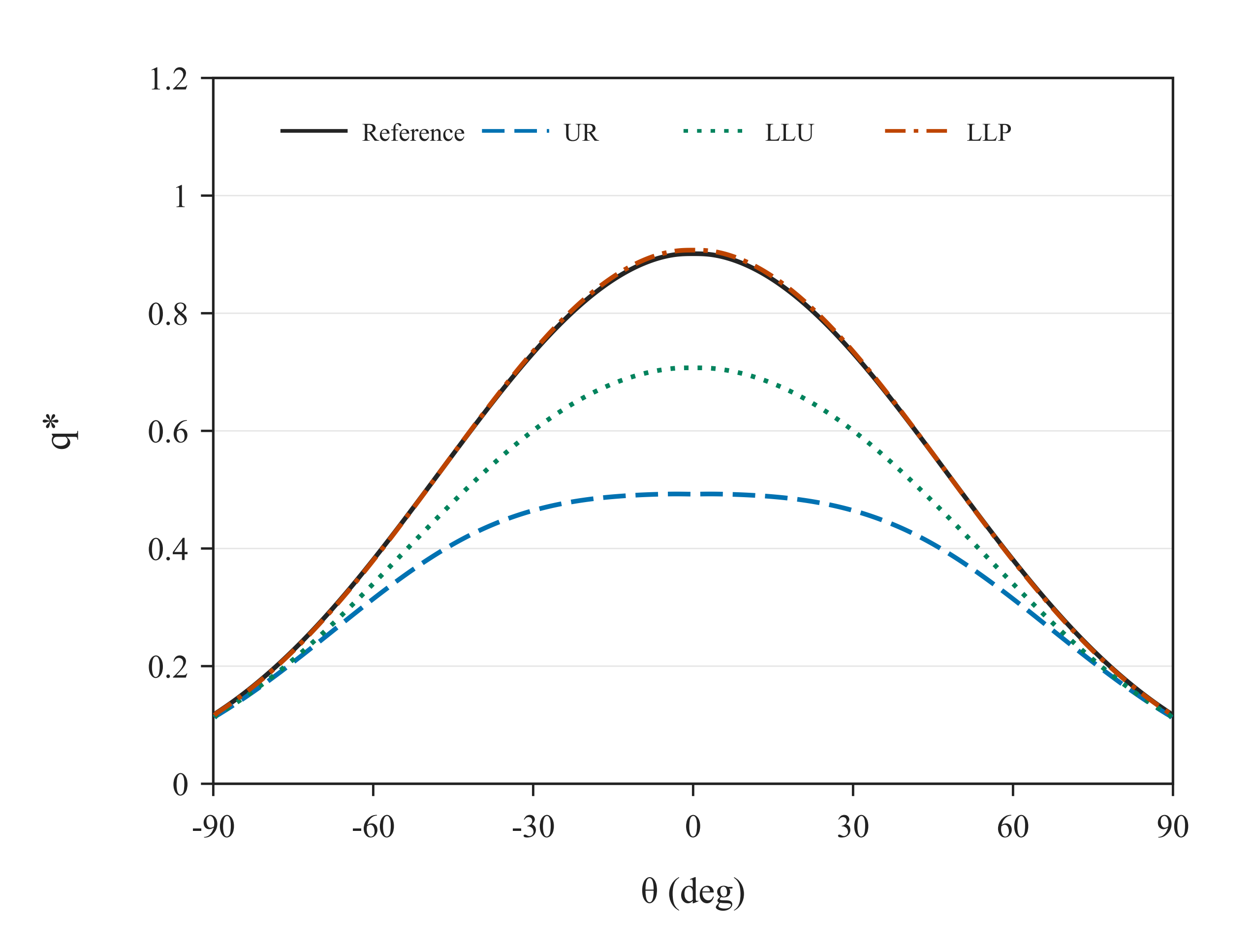}{GG, level 4}\hfill
		\panel{.32}{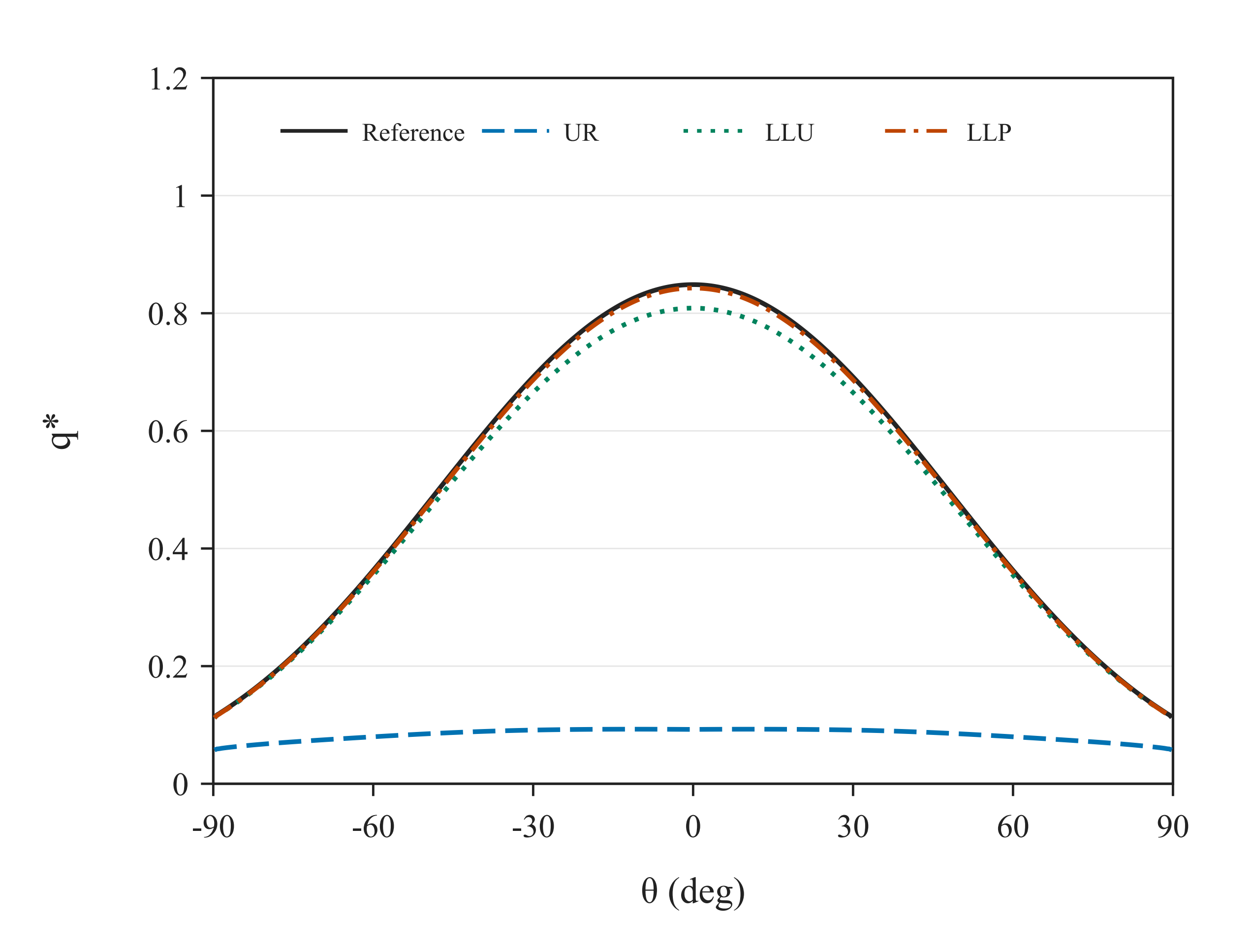}{GC, level 4}\hfill
		\panel{.32}{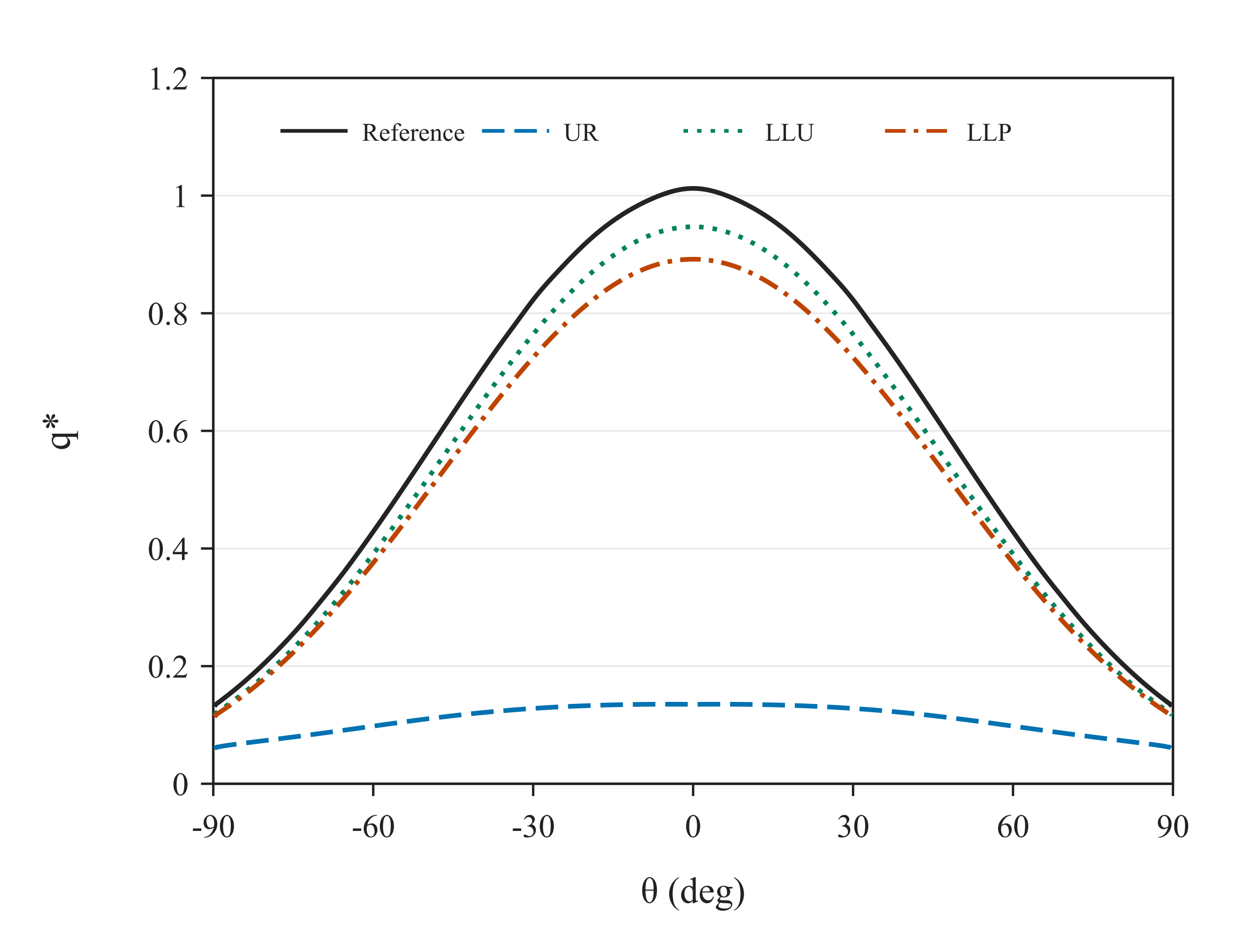}{POS, level 4}
		\caption{Wall heat flux on the three mesh families. Black solid denotes the scheme-specific reference, blue dashed denotes UR, green dotted denotes LLU and orange dash-dotted denotes LLP. The same axis limits are used in all panels.}
		\label{fig:qmesh}
	\end{figure}

	Figure~\ref{fig:qmesh} compares the heat-flux curves at the two design levels, and Table~\ref{tab:qmesh} gives the corresponding signed near-stagnation changes and full-curve differences. UR produces the largest full-curve change for every scheme and at both levels. The reduction is visible across the surface curve and is especially pronounced for the compact schemes. GG is less sensitive to UR in this case, although its level-4 response remains substantial.
	
	% Figure 6
	
	\begin{table}[!tb]
		\centering
		\caption{Heat-flux sensitivity relative to the reference of each scheme. Entries are percentages. $E_{\rm all}$ uses all 400 wall faces.}
		\label{tab:qmesh}
		\begin{tabular}{lrrrrrr}
			\toprule
			& \multicolumn{2}{c}{GG} & \multicolumn{2}{c}{GC} & \multicolumn{2}{c}{POS} \\
			\cmidrule(lr){2-3}\cmidrule(lr){4-5}\cmidrule(lr){6-7}
			Mesh & $\delta_s$ & $E_{\rm all}$ & $\delta_s$ & $E_{\rm all}$ & $\delta_s$ & $E_{\rm all}$ \\
			\midrule
			$\mathrm{UR}_2$ & -21.25 & 17.76 & -60.55 & 54.88 & -69.13 & 63.53 \\
			$\mathrm{LLU}_2$ & -16.04 & 13.76 & -4.64 & 4.21 & -7.87 & 8.33 \\
			$\mathrm{LLP}_2$ & 0.11 & 0.15 & -0.23 & 0.25 & -4.84 & 4.92 \\
			$\mathrm{UR}_4$ & -45.36 & 38.15 & -89.11 & 86.25 & -86.66 & 84.07 \\
			$\mathrm{LLU}_4$ & -21.53 & 18.59 & -4.71 & 4.01 & -6.45 & 6.99 \\
			$\mathrm{LLP}_4$ & 0.67 & 0.49 & -0.69 & 0.71 & -11.89 & 11.78 \\
			\bottomrule
		\end{tabular}
	\end{table}
	
	LLU is substantially less disruptive than UR at both levels. The retained interval count provides more sampling across the modified band despite similar first heights. Its response depends on the complete scheme and is not monotone between the two levels for every metric; Table~\ref{tab:qmesh} shows this explicitly.
	
	LLP leaves the GG and GC heat-flux curves close to their own references. Within POS, LLP gives the smallest full-curve difference at level 2, whereas LLU gives the smallest one at level 4. Thus retaining the outer part of the original spacing can be effective, but the response of POS also depends on the details of the innermost redistribution. A single first-cell height cannot account for the differences between UR, LLU and LLP.
	
	\subsection{Sensitivity of the reported pressure diagnostics}
	
\begin{figure}[!tb]
		\centering
		\panel{.32}{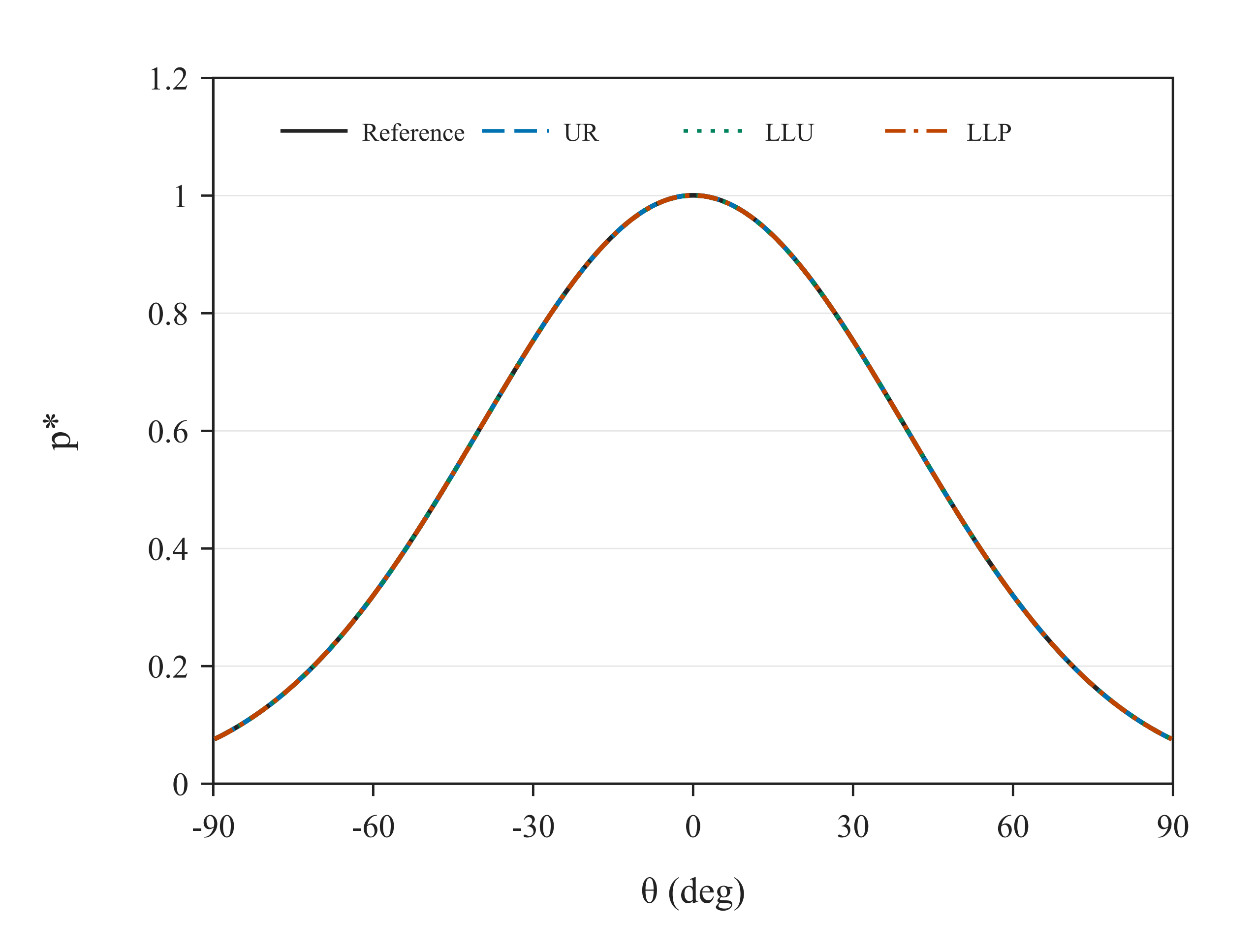}{GG}\hfill
		\panel{.32}{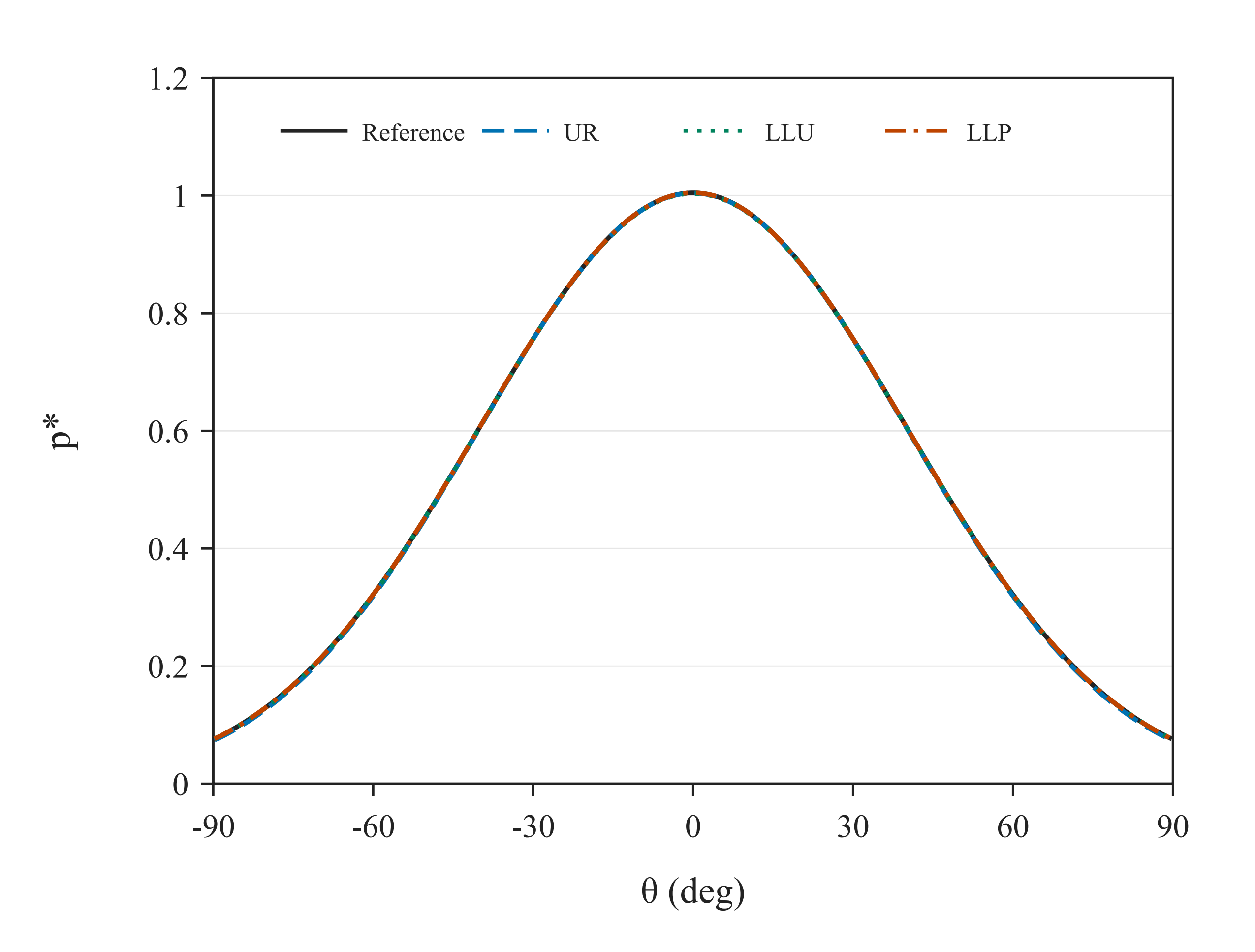}{GC}\hfill
		\panel{.32}{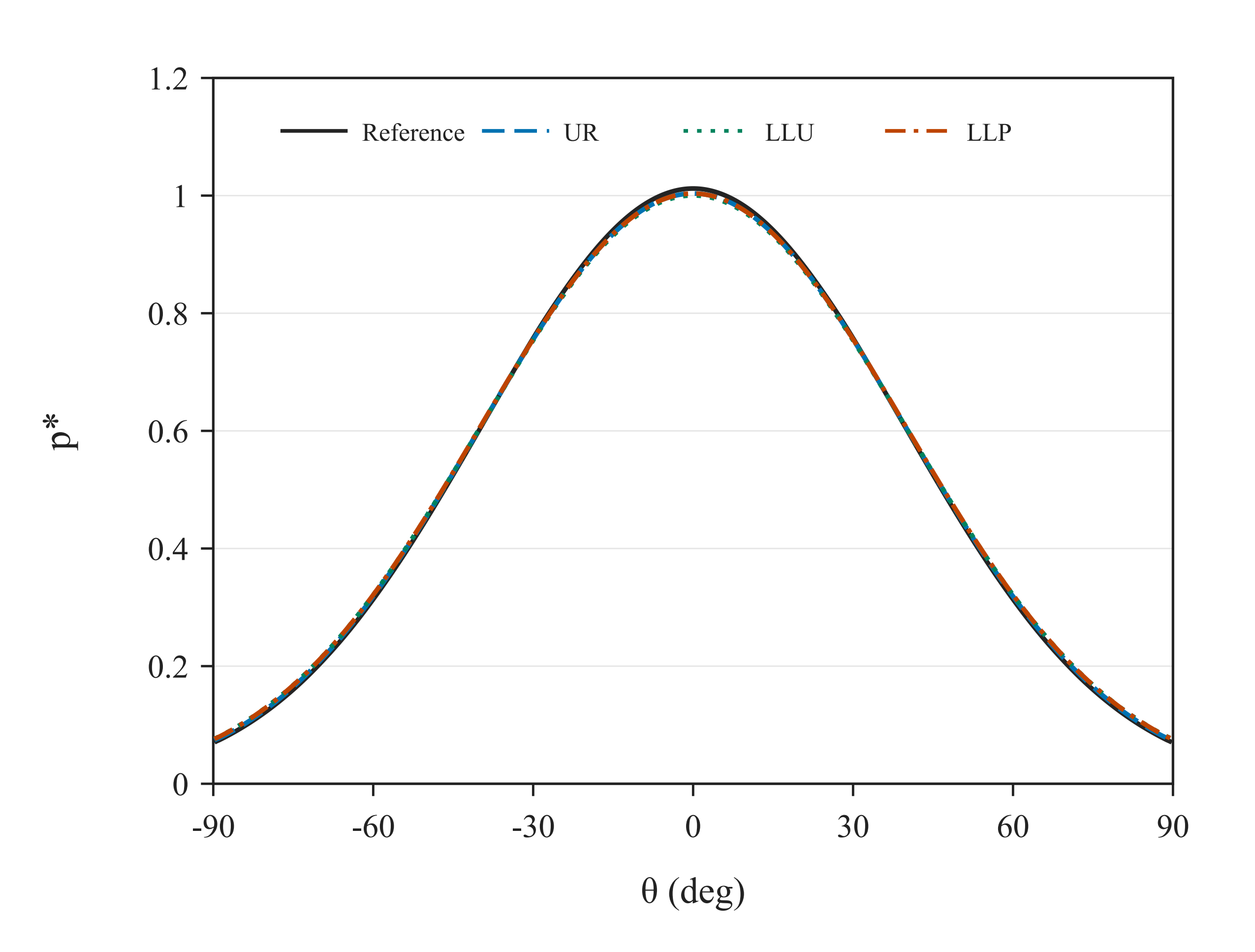}{POS}
		\caption{Reported wall-pressure diagnostics at mesh level 4. Curve styles follow Fig.~\ref{fig:qmesh}. The diagnostic is compared only with the reference of the same scheme.}
		\label{fig:pmesh}
	\end{figure}

	Figure~\ref{fig:pmesh} shows the reported pressure diagnostics on the reference and level-4 meshes. The curves are visually close on the full-angle scale. Table~\ref{tab:pmesh} gives differences from each scheme's own reference for both levels and distinguishes the full wall from the central sector. Across all 18 modified cases, the largest full-wall difference in a reported pressure diagnostic is 1.14\%, on POS--LLU4. For the three UR4 calculations, the full-wall differences are about 0.03\%, 0.29\% and 0.77\% for GG, GC and POS, respectively. These within-scheme changes are much smaller than the corresponding heat-flux responses. Since the pressure output is sampled differently in GG, GC and POS, neither these differences nor the visual proximity of the curves defines a common pressure observable for an interscheme accuracy comparison.
	
	% Figure 7
	
	\begin{table}[!tb]
		\centering
		\caption{Wall-pressure sensitivity relative to the reference of each scheme. Entries are percentages. $E_{15}$ uses $|\theta|\leq15^\circ$.}
		\label{tab:pmesh}
		\begin{tabular}{lrrrrrr}
			\toprule
			& \multicolumn{2}{c}{GG} & \multicolumn{2}{c}{GC} & \multicolumn{2}{c}{POS} \\
			\cmidrule(lr){2-3}\cmidrule(lr){4-5}\cmidrule(lr){6-7}
			Mesh & $E_{\rm all}$ & $E_{15}$ & $E_{\rm all}$ & $E_{15}$ & $E_{\rm all}$ & $E_{15}$ \\
			\midrule
			$\mathrm{UR}_2$ & 0.01 & 0.00 & 0.08 & 0.04 & 0.92 & 0.78 \\
			$\mathrm{LLU}_2$ & 0.02 & 0.02 & 0.03 & 0.04 & 1.00 & 0.85 \\
			$\mathrm{LLP}_2$ & 0.01 & 0.01 & 0.02 & 0.02 & 0.38 & 0.31 \\
			$\mathrm{UR}_4$ & 0.03 & 0.00 & 0.29 & 0.05 & 0.77 & 0.79 \\
			$\mathrm{LLU}_4$ & 0.03 & 0.03 & 0.08 & 0.10 & 1.14 & 1.10 \\
			$\mathrm{LLP}_4$ & 0.01 & 0.01 & 0.03 & 0.00 & 0.98 & 0.79 \\
			\bottomrule
		\end{tabular}
	\end{table}
	
	\subsection{Temperature structure on the stagnation line}
	
\begin{figure}[!tb]
		\centering
		\panel{.32}{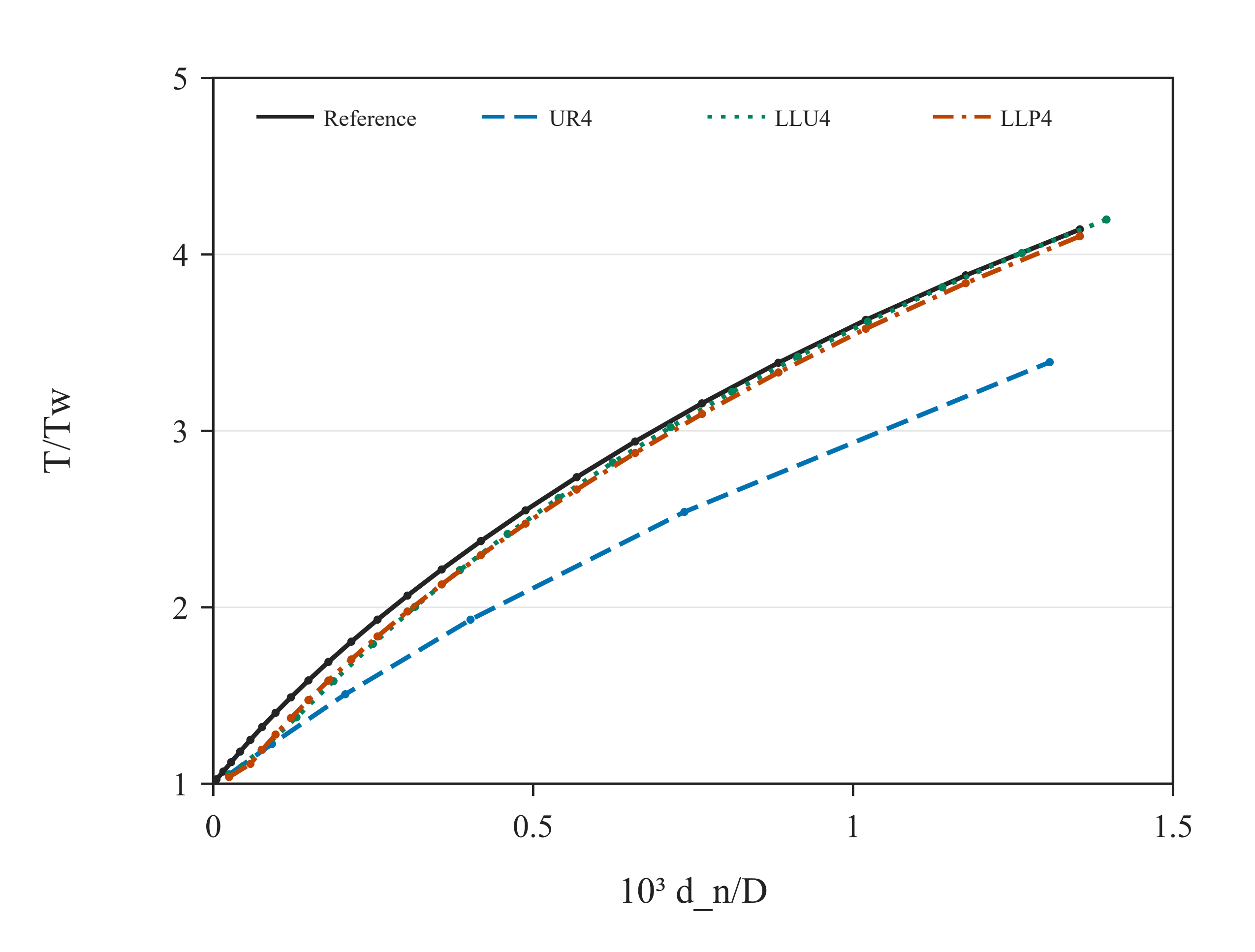}{GG}\hfill
		\panel{.32}{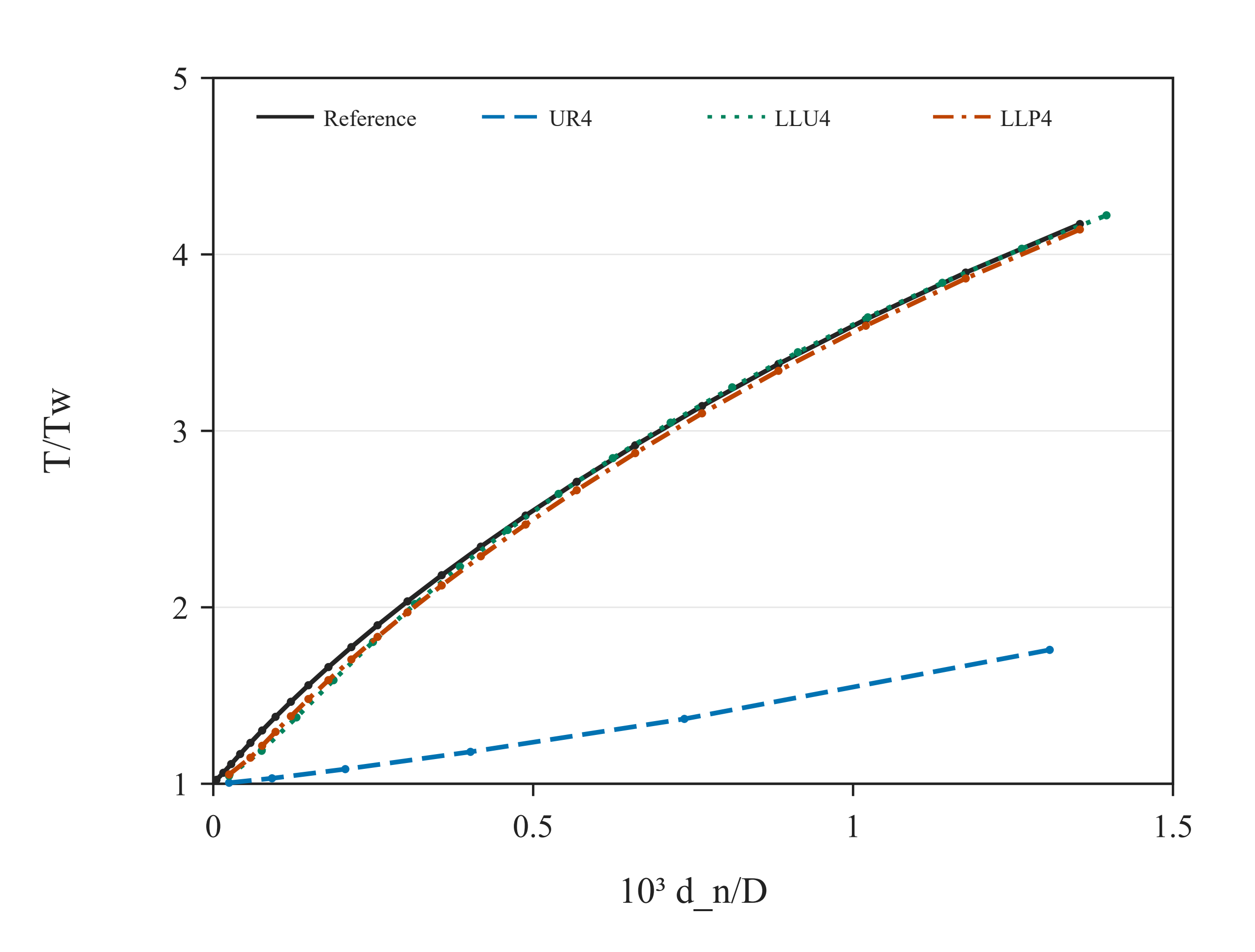}{GC}\hfill
		\panel{.32}{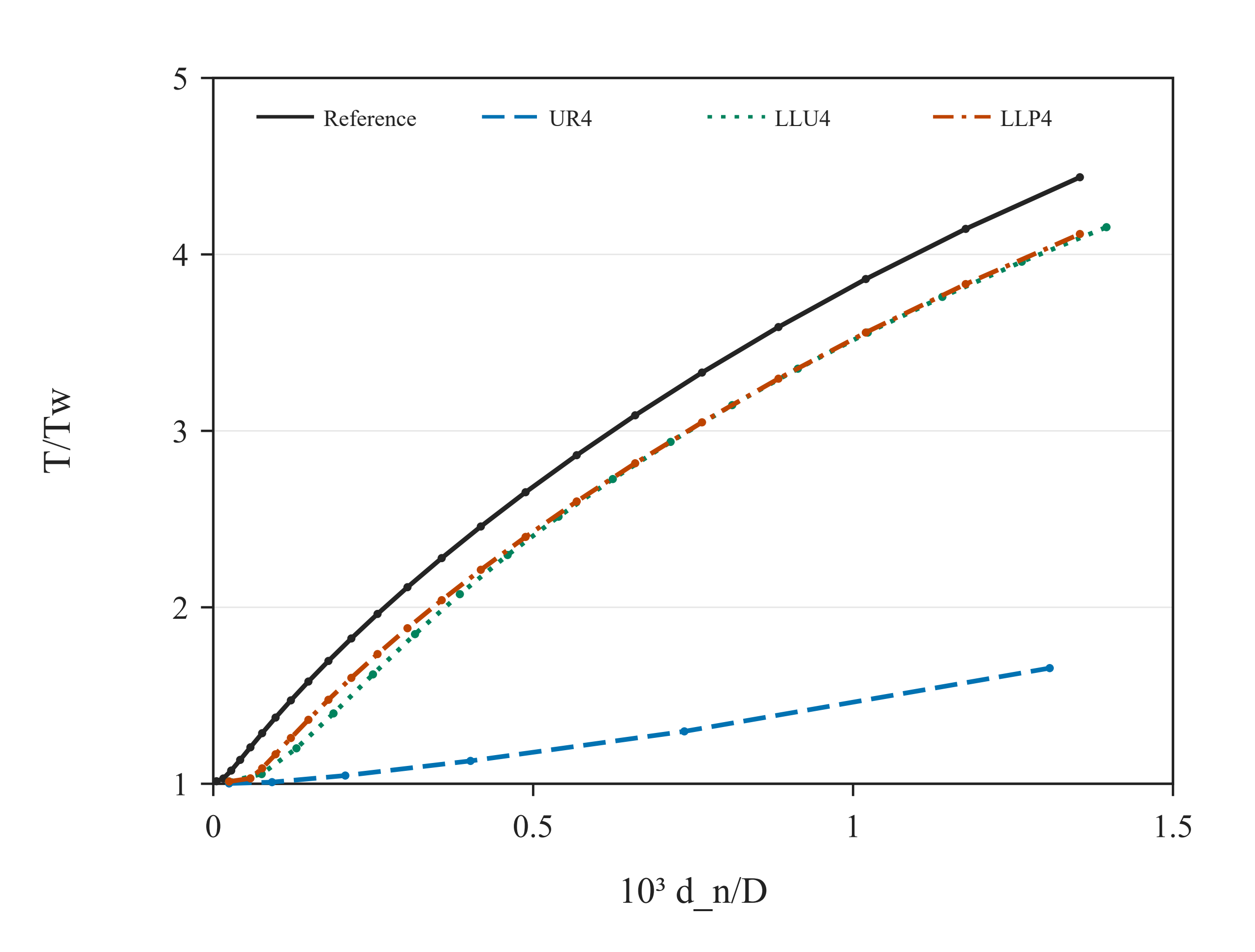}{POS}
		\caption{Local view of the stagnation-line temperature derived from stored cell-average states. Here $d_n$ is wall-normal distance and the plot covers $0\leq10^3d_n/D\leq1.5$. Markers give individual interval samples and lines only connect these samples. Each panel identifies the reference, UR4, LLU4 and LLP4 curves directly.}
		\label{fig:temperature}
	\end{figure}

	Figure~\ref{fig:temperature} enlarges the wall-adjacent portion of the temperature profiles. Primitive temperature is obtained from the stored conservative state in each cell using Eq.~\eqref{eq:state}. For each interval along $y=0$, the temperatures from the two cells adjacent to that interval are averaged and placed at the interval midpoint. Thus the markers are samples derived from cell averages; they are not temperatures reconstructed at the wall flux points. In particular, the first value is inside the first interval, and its distance from the wall changes with the mesh.
	
	% Figure 8
	
	In all three schemes the UR4 temperature rises more slowly over the displayed wall-adjacent region than in the corresponding reference calculation. The LLU4 and LLP4 cell-average profiles remain closer to the reference over much of this region. This is qualitatively consistent with the strong UR heat-flux reduction, but it does not identify the wall-point gradient used by the kinetic flux. The heat flux is obtained from separately stored wall output, while Fig.~\ref{fig:temperature} is postprocessed from cell-average flow fields. Since complete residual and heat-flux histories are not available for every case, the figure is not used to claim that the two plotted quantities were extracted from an independently verified identical convergence state.
	
	\subsection{Implications for the wall treatment}
	
	The comparison separates two aspects of the calculation. The point constraints specify wall data at the flux locations and determine the unlimited quadratic in a full-rank stencil. The mesh determines how much information about the thermal layer is available in the neighboring averages. The large UR responses of both GC and POS show that a compact quadratic representation, even with local wall constraints, remains sensitive to broad loss of wall-normal resolution. For this cylinder calculation, the smaller LLU and LLP changes are consistent with retaining more intervals across the band. At level 2, LLP gives the smallest POS full-curve deviation from its reference; at level 4, LLU gives the smallest such deviation. These statements concern the specified metric and this complete scheme; they do not imply a general ranking of mesh modifications or an isolated effect of the point constraints.
	
	The benchmarks concern the complete GG, GC and POS schemes, with their respective wall treatments and geometry representations. The local derivation in Section~\ref{sec:pointwise} explains what the second wall equation changes in the reconstruction. The numerical comparisons quantify how the three complete schemes respond to the specified mesh modifications. The small within-scheme variation of the reported pressure diagnostics and the larger heat-flux variations provide complementary but non-equivalent evidence about the computed flow.
	
	\section{Conclusions}
	\label{sec:conclusions}
	
	This work compared wall-heat-flux predictions from three complete gas-kinetic schemes for Mach~8.03 cylinder flow: the linear Green--Gauss scheme (GG), the compact ghost-cell scheme (GC), and the pointwise one-sided compact scheme (POS).  At the same 24 supplied heat-flux angles, GG has the smallest RMS difference on the reference, UR2, UR4, LLP2, and LLP4 meshes; POS is closest on LLU2 and LLU4; and GC is not the lowest-RMS scheme on any tested mesh.  Thus, no treatment is uniformly superior.  For the reference, UR, and LLP families, GG is the more robust practical choice among the compared schemes, while POS is preferable under the fixed-count node redistribution of the LLU family.
	
	The GG result on the reference, UR, and LLP meshes is consistent with the large aspect ratios of the near-wall cells.  GC and POS solve a local quadratic reconstruction system involving cell averages, gradients, and wall-condition rows.  Although reference-coordinate reconstruction scales the polynomial basis, physical moments, wall normals, and recovered gradients retain geometric sensitivity.  GG instead uses a face-based linear gradient and avoids this constrained quadratic system, making it less exposed to this possible conditioning effect on the present stretched cells.  This explanation is plausible rather than proven because local condition numbers and wall-constraint residuals were not saved.  The LLU results show that wall-normal node placement can change the preferred treatment.
	
	Mesh coarsening is a secondary observation.  Uniform reduction produces the largest full-curve heat-flux response for all three schemes, whereas the pressure curves change much less.  The comparison concerns complete schemes, not an isolated effect of the POS point constraints, since reconstruction, wall data, geometry representation, and pressure sampling differ among the implementations.  A controlled common-geometry comparison is needed before generalizing these mesh-family-specific conclusions.

	\appendix
	\section{Boundary rows and wall energy flux}
	
	\subsection{Finite-volume moments on the illustrative stencil}
	
	On the unit-square stencil of Section~\ref{sec:assembly}, the three neighbor mean rows are
	\begin{equation}
		\bm D=
		\begin{bmatrix}
			-1&0&1&0&0\\
			1&0&1&0&0\\
			0&1&0&0&1
		\end{bmatrix}.
	\end{equation}
	For example, the mean of $\xi^2$ over a unit square centered at $\xi_c$ is $\xi_c^2+1/12$, while the target mean is $1/12$. The mean of $\xi\eta$ is the product of the center coordinates and vanishes in all four cells. Writing $g=1/(2\sqrt3)$, the value and outward normal-derivative rows on the wall are
	\begin{align}
		\bm c_{V,\pm}&=(\pm g,-1/2,0,\mp g/2,1/6),\\
		\bm c_{N,\pm}&=(0,-1,0,\mp g,1).
	\end{align}
	The geometry is chosen for the algebraic check. In the curved reconstruction, the moment and normal rows are evaluated on the computational cells. The derivative row follows directly from
	\[
	\nn_k^{\mathsf T}\nabla_{\xx}P_Q
	= (\bm J^{-1}\nn_k)^{\mathsf T}\nabla_{\bm r}P_Q .
	\]
	This identity applies to a skew affine coordinate map as well as to the aligned example.
	
	\subsection{Normal mass balance for the two time coefficients}
	
	Let $B_w=\sqrt{R_gT_w/(2\pi)}$. The outgoing unit-density Maxwellian has normal half-space moment $-B_w$ because its one-dimensional normal marginal is a centered Gaussian with variance $R_gT_w$. The constant and linear outgoing mass moments are consequently $-B_w\rho_R$ and $-B_w\rho_RA_R$. Matching them to the negatives of the incident moments gives
	\begin{equation}
		m_0-B_w\rho_R=0,
		\qquad m_1-B_w\rho_RA_R=0 .
	\end{equation}
	The cancellation concerns the mass component of the kinetic flux. The corresponding energy moment is calculated with the kinetic-energy invariant and is generally nonzero. This is how impermeability and energy exchange coexist in the isothermal flux. The reconstructed density target in Eq.~\eqref{eq:density} enters through the incident state, while $\rho_R$ supplies the outgoing half-space distribution.

\end{document}